\documentclass[useAMS,usenatbib]{mn2e} 

\usepackage{graphicx}
\usepackage{rotating}
\DeclareGraphicsExtensions{.eps,.ps,.eps.gz,.ps.gz,.eps.Z}
\title[Dust and Metal Column Densities in GRB Host Galaxies]{Dust and Metal Column Densities in Gamma-Ray Burst Host Galaxies}

\author[P.~Schady et al.]{P.~Schady$^1$, M.J.~Page$^1$, S.R.~Oates$^1$, M.~Still$^1$, M.De~Pasquale$^1$, T.~Dwelly$^2$, \and N.P.M.~Kuin$^1$, S.T.~Holland$^{3,4,5}$, F.E.~Marshall$^{3}$ and P.W.A.~Roming$^6$\\
$^1$ The UCL Mullard Space Science Laboratory, Holmbury St Mary, Dorking, Surrey, RH5 6NT, UK.\\
$^2$ School of Physics and Astronomy, University of Southampton, Highfield, Southampton SO17 1BJ, UK\\
$^3$ Astrophysics Science Division, Code 660.1, 8800 Greenbelt Road, GSFC, Greenbelt, MD 20771, USA\\
$^4$ Universities Space Research Association, 10211 Wincopin Circle, Suite 500, Columbia, MD 21044, USA\\
$^5$ CRESST, Code 668.8, 8800 Greenbelt Road, GSFC, Greenbelt, MD 20771, USA\\
$^6$ Department of Astronomy and Astrophysics, PSU, 525 Davey Laboratory, University Park, PA 16802, USA
}

\date{Accepted: Received: }

\begin{document}

\newcommand\eg{e.g. } 
\newcommand\ie{i.e. } 
\newcommand\swift{{\it Swift}} 
\newcommand\sqrcm{cm$^{2}$} 
\newcommand\invsqrcm{cm$^{-2}$} 
\newcommand\flu{$\mathrm{erg}~\mathrm{cm}^{-2}$}
\newcommand\flux{$\mathrm{erg}~\mathrm{s}^{-1}$}
\newcommand\cps{$\mathrm{counts}~\mathrm{s}^{-1}$}
\newcommand\nhx{$N_{H,X}$}
\newcommand\nh{$N_{HI}$}
\newcommand\nhinf{$N_{infH}$}
\newcommand\av{$A_V$}
\newcommand\Rv{$R_V$}
\newcommand\Eiso{$E_{iso}$}
\newcommand{\T}{T$_{90}$}
\def\lesssim{\mathrel{\hbox{\rlap{\hbox{\lower4pt\hbox{$\sim$}}}\hbox{$<$}}}}
\def\gtrsim{\mathrel{\hbox{\rlap{\hbox{\lower4pt\hbox{$\sim$}}}\hbox{$>$}}}}

\maketitle 

\begin{abstract}
In this paper we present the results from the analysis of a sample of 28 gamma-ray burst (GRB) afterglow spectral energy distributions, spanning the X-ray through to near-infrared wavelengths. This is the largest sample of GRB afterglow spectral energy distributions thus far studied, providing a strong handle on the optical depth distribution of soft X-ray absorption and dust-extinction systems in GRB host galaxies. We detect an absorption system within the GRB host galaxy in 79\% of the sample, and an extinction system in 71\% of the sample, and find the Small Magellanic Cloud (SMC) extinction law to provide an acceptable fit to the host galaxy extinction profile for the majority of cases, consistent with previous findings. The range in the soft X-ray absorption to dust-extinction ratio, \nhx/\av, in GRB host galaxies spans almost two orders of magnitude, and the typical ratios are significantly larger than those of the Magellanic Clouds or Milky Way. Although dust destruction could be a cause, at least in part, for the large \nhx/\av\ ratios, the good fit provided by the SMC extinction law for the majority of our sample suggests that there is an abundance of small dust grains in the GRB environment, which we would expect to have been destroyed if dust destruction were responsible for the large \nhx/\av\ ratios. Instead, our analysis suggests that the distribution of \nhx/\av\ in GRB host galaxies may be mostly intrinsic to these galaxies, and this is further substantiated by evidence for a strong negative correlation between \nhx/\av\ and metallicity for a subsample of GRB hosts with known metallicity. Furthermore, we find the \nhx/\av\ ratio and metallicity for this subsample of GRBs to be comparable to the relation found in other more metal-rich galaxies.
\end{abstract}


\section{Introduction}
To unravel the properties of gamma-ray burst (GRB) progenitors and the fundamental conditions required within a galaxy to form a GRB, an understanding of the GRB circumburst and host galaxy environment is essential. The faint optical magnitudes and large distances of GRB hosts limit the amount of information obtained from host galaxy observations, and thus studying the spectral properties of the afterglow provide the most direct and sensitive method of probing the surrounding environments of GRBs.

Both optical and X-ray spectroscopic observations have provided detailed information on the metal abundances and column densities in GRB local environments \citep[e.g.][]{pcd+07,sff03, sf04}, and broadband analysis of the GRB afterglow spectral energy distribution (SED) allows the host galaxy dust extinction curve to be well modelled, and thus provides a measure of the host visual extinction. \citet{gw01} combined these two techniques to compare the soft X-ray absorption with the visual dust extinction in the local environment of a sample of eight GRBs. In their analysis, they found that whereas the distribution of equivalent neutral hydrogen column density within GRB host galaxies was comparable to that observed in Galactic molecular clouds, the measured host galaxy visual extinction was 10--100 times smaller than expected for GRBs embedded in Galactic-like molecular clouds. \citet{sfa+04} expanded on this work, and found the gas column density to dust extinction ratio to not only be larger than that of the Milky Way, but also to be around an order of magnitude larger than that of the gas rich Large and Small Magellanic Clouds (LMC and SMC, respectively). In both \citet{gw01} and \citet{sfa+04} the large column density to visual extinction ratios measured in GRB local environments was taken to be evidence of dust destruction by the GRB, causing the visual extinction to decrease. In addition to this, \citet{sfa+04} also found that the optical to near-infrared (NIR) GRB afterglow data showed little evidence of the strong 2175~\AA\ absorption feature present in the Milky Way extinction law, and less prominently in the LMC. Instead, they found the mean SMC or the \citet{dks+94} starburst galaxy dust extinction law to provide the best fit to their sample of GRB SEDs, both of which have no 2175~\AA\ absorption feature. These results are supported by the more recent work done by \citet{kkz06} and \cite{sww+07}. In each of these cases the mean SMC, LMC and Galactic extinction curves were assumed. However, a range in the total-to-selective extinction, $R_V=A_V/E(B-V)$, which relates the reddening to extinction, and 2175~\AA\ bump strength is observed along different lines-of-sight through the Milky Way \citep{ccm89} and the Magellanic Clouds \citep{gcm+03}, and observations of higher redshift supernovae and quasars ($z > 5$) also suggest differences between the dust extinction properties in higher redshift galaxies and the local universe \citep[e.g.][]{mso+04,td01}. However, the degeneracy that exists between the best-fit GRB spectral index and the host galaxy's total-to-selective extinction means that to accurately determine the host galaxy's extinction law properties, good quality, broadband data are needed, preferentially stretching out into the negligibly extinguished far infrared (FIR) wavelength bands.

In the current era of \swift\ and rapid-response ground-based telescopes, prompt arcsecond GRB positions have provided a wealth of high quality, early-time X-ray, ultraviolet (UV), optical and NIR data. Accurate soft X-ray absorption measurements are now available for the large fraction of GRBs \citep{crc+06,bk07,gnv+07,smp+07,ebp+09}, and well-sampled, high signal-to noise SEDs are providing strong constraints on the best-fit extinction law models \citep[e.g.][]{pbb+08}. There have now been some examples of GRB host galaxies with the 2175~\AA\ absorption feature (e.g. GRB~070802, El{\'{\i}}asd{\'o}ttir et al. 2009, Kr{\"u}hler et al. 2008; GRB~080607, Prochaska et al. 2009), as well as GRB host galaxies with \Rv\ values larger than the mean SMC, LMC and MW values \citep[e.g.][]{pbb+08}, a possible indicator of grey dust, as suggested to be present in some GRB host galaxies \citep[e.g.][]{sff03}. However, such analysis on the detailed properties of GRB extinction curves are typically still only possible for a handful of well-sampled, bright GRBs (e.g. GRB050525A, Heng et al. 2008; GRB~061126, Perley et al. 2009; GRB~070802, El{\'{\i}}asd{\'o}ttir et al. 2009, Kr{\"u}hler et al. 2008; GRB~080607, Prochaska et al. 2009).

In \citet{smp+07} we used X-ray and UV/optical simultaneous observations taken with the X-ray Telescope \citep[XRT;][]{bhn+05} and Ultraviolet and Optical Telescope \citep[UVOT;][]{rkm+05} onboard \swift\ \citep{gcg+04} to analyse the SEDs for a sample of 7 GRBs. The dust extinction in the GRB host galaxies was modelled on the mean SMC, the LMC and the Milky Way extinction curves using the parameterisations given in \citet{pei92}, which cover a range in 2175~\AA\ bump strengths and \Rv\ values. The SMC and LMC extinction curves were found to provide the best-fit model for the majority of the sample, in agreement with previous studies \citep[e.g.][]{sfa+04,kkz06,sww+07}. However, we also found that, although the gas-to-dust ratio in \swift\ GRB host galaxies were typically larger than those of the Milky Way and Magellanic Clouds, the weighted mean was within 90\% confidence of the Magellanic Clouds and Milky Way X-ray absorption to optical extinction ratios. 

In this paper we aim to further our previous work, using a larger sample of 28 GRBs, and increasing the wavelength range of the afterglow SEDs to better constrain the absorption and extinction within the GRB host galaxy. In \citet{smp+07} \swift\ data alone were used to produce the SEDs, and UVOT data with a rest-frame wavelength $\lambda < 1215$~\AA\ were not included in the SED fits in order to avoid the absorption caused by the Lyman forest being confused for dust-extinction. In this paper we now model the absorption resulting from the Lyman forest such that all rest frame UV data redward of the Lyman edge is included in our spectral analysis. Furthermore, we also include additional ground-based NIR data if available, further increasing the spectral range of the SEDs and the degrees of freedom of the spectral fits. This provides better sampled SEDs and extends the redshift range within our sample, which was previously restricted to $z<1.7$ to ensure that the SED modelling was sufficiently well constrained within the optical wavelength range. 

In $\S$\ref{sec:analysis} we present the new, extended GRB sample and describe the X-ray, UV/optical and NIR data reduction and analysis, and in $\S$\ref{sec:model} we describe the models used to fit the data. We present the results of our spectral modelling in $\S$\ref{sec:results} followed by an analysis of the possible selection effects and systematic biases that may be present in our work in $\S$\ref{sec:seleffs}. A discussion on the implications of our findings is presented in $\S$\ref{sec:disc}, and our conclusions are summarised in $\S$\ref{sec:cons}. Throughout the paper temporal and spectral indices, $\alpha$ and $\beta$, respectively, are denoted such that $F(\nu,t)\propto \nu^{-\beta}t^{-\alpha}$, and all errors are $1\sigma$ unless specified otherwise.

\begin{table*}
\caption{Table listing the 28 GRBs in our sample with their redshift, Galactic column density and visual extinction in the line-of-sight to the GRB, the corresponding SED epoch, the UV, optical, and NIR band passes included in the GRB afterglow SED, and the rest frame coverage of the SED.\label{tab:sample}}
\begin{tabular}{@{}lllllll}
\hline
GRB & z & $N_{H,X}$(Gal) &  $A_V$(Gal) & Epoch & UV/optical/NIR bandpasses & Restframe Band \\
 & & ($10^{21}$~\invsqrcm) & & (s) & & Coverage (\AA) \\
\hline\hline
050318 & 1.44$^a$ & 0.28 & 0.05 & T$^\dag$+3600 & v,b,u & 1260--2400 \\
050319 & 3.24$^b$ & 0.11 & 0.03 & T+20,000 & $I^{1,2},R^{1-3}$,v,b & 920--2090 \\
050525A & 0.606$^c$ & 0.91 & 0.29 & T+20,000 & $K^{4,5},H^4,J^{4,6},I^{4,8},R^{4,9,10}$,v,b,u,w1,m2,w2 & 1000-14540 \\
050730 & 3.968$^d$ & 0.30 & 0.16 & T+10,000 & $K^{11},J^{11},I^{11},R^{11}$,v,b & 780--4700 \\
050802 & 1.71$^e$ & 0.18 & 0.06 & T+20,000 & $I^{12},R^{12}$,v,b,u,w1,m2,w2 & 590--3270 \\
050820A & 2.6147$^f$ & 0.47 & 0.14& T+10,000 & $J^{13},z^{14},I^{14},R^{14},g^{14}$,v,b,u,w1 & 620--3710 \\
050922C & 2.198$^g$ & 0.54  & 0.32 & T+20,000 & $R^{15-20}$,v,b,u,w1,m2 & 620--2360 \\
051109A & 2.346$^h$ & 1.61 & 0.59 & T+5000 & $K^{21},H^{21},J^{21},I^{22},R^{23,24}$,v,b,u,w1 & 670--6980 \\
060124 & 2.296$^i$ & 0.92 & 0.42 & T+100,000 & $I^{25},R^{25}$,v,b & 1180--2690 \\
060206 & 4.048$^j$ & 0.09 & 0.04 & T+10,000 & $K^{26},H^{26},J^{26},R^{27-29}$,v,b & 770--4630\\
060418 & 1.49$^k$ & 0.92 & 0.69 & T+5000 & $K^{30},H^{30},J^{30},z^{30,31},I^{32},R^{33}$,v,b,u,w1,m2 & 800--9380 \\
060502A & 1.51$^l$ & 0.30 & 0.10 & T+5000 & $R^{34}$,v,b,u,w1 & 900--3010 \\
060512 & 0.4428$^m$ & 0.14 & 0.05 & T+10,000 & $Ks^{35},J^{36},R^{37,38}$,v,b,u & 2130--16180 \\
060526 & 3.21$^n$ & 0.55 & 0.21 & T+20,000 & $J^{39},I^{39,40},R^{40-47}$,v,b & 930--3180 \\
060605 & 3.711$^o$ & 0.51 & 0.15 & T+10,000 & $R^{48-53}$,v,b & 830--1600 \\
060607A & 3.082$^p$ & 0.27 & 0.09 & T+10,000 & $H^{30},i^{54},r^{54},g^{54}$,v,b,u & 750--4200 \\
060714 & 2.71$^q$ & 0.61 & 0.24 & T+5000 & $R^{55}$,v,b & 1050--2040 \\
060729 & 0.54$^r$ & 0.48 & 0.17 & T+70,000 & $R^{56}$,v,b,u,w1,m2,w2 & 1040--4900 \\
060904B & 0.703$^s$ & 1.21 & 0.53 & T+5000 & $K^{57},J^{57},I^{57 },R^{58,59}$,v,b,u,w1,m2,w2 & 940--13710 \\
060908 & 2.43$^t$ & 0.27 & 0.09 & T+5000 & $R^{60,61}$,v,b,u,w1 & 660--2200 \\
060912 & 0.937$^u$ & 0.42 & 0.16& T+1500 & v,b,u,w1,m2 & 1030--3020 \\
061007 & 1.262$^v$ & 0.21 & 0.06 & T+600 & $i^{62},R^{62}$,v,b,u,w1,m2,w2 & 710--3850 \\
061121 & 1.314$^w$ & 0.51 & 0.14 & T+10,000 & $I^{62-64},R^{65}$,v,b,u,w1,m2,w2 & 690--3830 \\
061126 & 1.159$^x$ & 1.00 & 0.56 & T+2000 & $K^{66},J^{66},I^{66,67},R^{66,68,69}$,v,b,u,w1,m2 & 920--10820 \\
070110 & 2.352$^y$ & 0.18  & 0.04 & T+10,000 & $R^{70}$,v,b,u & 920--2250 \\
070318 & 0.836$^z$ & 0.25 & 0.05 & T+1500 & v,b,u,w1,m2,w2 & 870--3190 \\
070411 & 2.954$^\ddag$ & 2.63 & 0.88 & T+500 & $R^{71,72}$,v,b & 990--1900 \\
070529 & 2.4996$^\S$ & 1.90 & 0.93 & T+600 & v,b,u,w1 & 640--1670 \\
\hline
\end{tabular}
 \begin{list}{}{}
 \item[]
 $^a$ \citet{bm05}; 
 $^b$ \citet{fhj+05};
 $^c$ \citet{fcb+05};
 $^d$ \citet{sve+05};
 $^e$ \citet{fsj+05};
 $^f$ \citet{lve+05};
 $^g$ \citet{jfp+05a};
 $^h$ \citet{qhr+05};
 $^i$ \citet{pft+06};
 $^j$ \citet{pwf+06};
 $^k$ \citet{dfp+06}
  $^l$ \citet{cpf+06};
 $^m$ \citet{bfk+06};
 $^n$ \citet{bg06};
  $^o$ \citet{fkk+06};
 $^p$ \citet{lvs+06};
 $^q$ \citet{jvf+06d};
 $^r$ \citet{tlj+06};
 $^s$ \citet{fdm+06};
 $^t$ \citet{rjt+06};
 $^u$ \citet{jlc+06c};
 $^v$ \citet{jft+06b};
 $^w$ \citet{bpc06};
 $^x$ \citet{pbb+08};
 $^y$ \citet{jmf+07};
 $^z$ \citet{jfa+07};
$^\ddag$ \citet{jmt+07};
 $^\S$ \citet{bfc07}
 \item[]$^\dag$ T is time at which the BAT triggered on the GRB
 \item[]
$^1$ \citet{huk+07};
$^2$ \citet{krm07};
$^3$ \citet{wvw+05};
$^4$ \citet{cb05};
$^5$ \citet{rg05};
$^6$ \citet{fhs+05}; 
$^7$ \citet{ffc+05};
$^8$ \citet{ytk05};
$^9$ \citet{hhg+05};
$^{10}$ \citet{mbs05a},
$^{11}$ \citet{pcm+06};
$^{12}$ \citet{pes+05};
$^{13}$ \citet{meb+05};
$^{14}$ \citet{ckh+06};
$^{15}$ \citet{dp05}; 
$^{16}$ \citet{jpt+05b};
$^{17}$ \citet{ap05};
$^{18}$ \citet{hkh+05};
$^{19}$ \citet{pmm+05};
$^{20}$ \citet{dpf+05}; 
$^{21}$ \citet{bbs+05};
$^{22}$ \citet{tor05};
$^{23}$ \citet{mwp+05};
$^{24}$ \citet{juc+05};
$^{25}$ \citet{mbs+07}; 
$^{26}$ \citet{apb06};
$^{27}$ \citet{cvw+07};
$^{28}$ \citet{sdp+07};
$^{29}$ \citet{wvw+06};
$^{30}$ \citet{mvm+07}; 
$^{31}$ \citet{nif+06}; 
$^{32}$ \citet{cob06a}; 
$^{33}$ \citet{kop06};
$^{34}$ \citet{cof06}, 
$^{35}$ \citet{hlm+06}; 
$^{36}$ \citet{sdp+06}; 
$^{37}$ \citet{cen06a};
$^{38}$ \citet{mil06};
$^{39}$ \citet{cob06b}; 
$^{40}$ \citet{tgb+06}; 
$^{41}$ \citet{bgv+06};
$^{42}$ \citet{cig+06},
$^{43}$ \citet{dhm+07}; 
$^{44}$ \citet{gtb+06};
$^{45}$ \citet{kbs+06a};
$^{46}$ \citet{md06};
$^{47}$ \citet{rpi+06}
$^{48}$ \citet{kg06};
$^{49}$ \citet{ksa+06c};
$^{50}$ \citet{ksa+06b};
$^{51}$ \citet{mfm+06};
$^{52}$ \citet{sap06};
$^{53}$ \citet{zqw+06};
$^{54}$ \citet{nrc+09}; 
$^{55}$ \citet{api06};
$^{56}$ \citet{qr06}; 
$^{57}$ \citet{cb06};
$^{58}$ \citet{pks+06}; 
$^{59}$ \citet{skv06}; 
$^{60}$ \citet{act+06};
$^{61}$ \citet{wtr06};
$^{62}$ \citet{mmg+07}; 
$^{62}$ \citet{cen06b};
$^{63}$ \citet{cob06c};
$^{64}$ \citet{tor06a};
$^{65}$ \citet{uau06};
$^{66}$ \citet{pbb+08};
$^{67}$ \citet{tor06b};
$^{68}$ \citet{smg+06};
$^{69}$ \citet{wm06};
$^{70}$ \citet{mjv07};
$^{71}$ \citet{msd07};
$^{72}$\citet{klk+07}
\end{list}
\end{table*}

\section{Data Reduction and Analysis}\label{sec:analysis}
The selection criteria for our sample is that the GRB must be long (\ie\ \T$>2$~s, where \T\ is the time-interval over which 90\% of the high-energy radiation ($\gtrsim 15$~keV) is emitted), it must have a spectroscopic redshift measurement, have been observed by the XRT and UVOT within an hour of the prompt emission, have a peak UVOT v-band magnitude v $\leq 19$, and be detected by the XRT and in at least three UV--IR filters (UVOT and/or ground based). The final requirement is needed in order to provide sufficient constraints for spectral fitting. A total of 28 \swift\ GRBs satisfied our selection criteria up to and including GRB~070529. By requiring that the GRBs in our sample have both a spectroscopic redshift measurement and UVOT v-band magnitude v $\leq 19$ we are introducing a bias against highly extinguished GRBs, that occur in very dusty regions of their host galaxy and/or along a line-of-sight with high foreground extinction. A number of previous studies have already shown that subsamples of GRBs with spectroscopic redshifts are biased against high obscuration (e.g. Fiore et al. 2007; Fynbo et al. 2009). Further to this, there is also a selection effect in the redshift distribution that biases against certain redshift ranges that have few prominent absorption lines in the observer frame optical bandpass, thus making it difficult to acquire an accurate spectroscopic redshift measurement. However, there is currently no evidence to suggest that there is a strong redshift dependence on the environmental conditions of GRB host galaxies, and therefore, for the purposes of this paper, where we are primarily interested in studying the dust and metal contents in the environments of GRBs, this selection effect in the GRB redshift distribution should not degrade our results. Due to the independence between foreground and host galaxy extinction, the bias against high foreground extinction should also not have any impact on our results on the GRB host extinction and absorption properties, and we therefore need to only worry about the selective effects introduced by large host galaxy extinction, which we explore in detail in section~\ref{ssec:seleffects}.

In order to measure the level of host galaxy dust-extinction and absorption in the GRB optical, UV and X-ray afterglows, we produced an SED at a single epoch for each of the 28 \swift\ detected GRBs in our sample, where the SED epoch was GRB dependent. All \swift\ data used to produce these SEDs were taken from the UK \swift\ data archive\footnotemark[1]. 
\footnotetext[1]{http://www.swift.ac.uk/swift\_portal/}
NIR data reported in refereed papers and GCNs were used to extend the afterglow SED to longer wavelengths, where preference was given to photometry from refereed journals. Furthermore, unlike in \citet{smp+07}, UVOT data with rest frame wavelengths $\lambda<1215$~\AA\ were also used. Absorption at these wavelengths caused by the Lyman forest was modelled using the work described in \citet{mad95}, which provides a statistical estimate of the number of intervening absorption systems in the line-of-sight as a function of redshift and column density, and thus opacity of the Lyman forest as a function of wavelength.

The epoch of the SED was chosen to minimise the total amount of interpolation required for each UV, optical and NIR photometric data point used in the SED. Since XRT and UVOT observations are taken simultaneously, this condition also limited the amount of interpolation required in the X-ray band, in which the GRB afterglow is typically detected for longer than in the UVOT \citep[e.g.][]{ops+09,ebp+09}. A further condition on the selected epoch of the afterglow SED was that there could not be any apparent spectral evolution in either the UVOT or XRT energy ranges during the interval used for photometric interpolation, as is sometimes observed during the early-time steep decay phase of the X-ray light curve \citep{nkg+06}, during flares \citep{fmr+07}, or in the presence of a supernova component \citep[\eg\ GRB~060218,][]{cmb+06}. The 28 GRBs in the sample are listed in Table~\ref{tab:sample}, together with their spectroscopic redshifts, the Galactic hydrogen column density and visual extinction in the line-of-sight to the GRB, the epoch of the SED, the UVOT and ground-based filters used in the SED, and the rest frame wavelength coverage.

\subsection{UVOT and Ground-Based Data}\label{ssec:opt}
The UVOT contains three optical and three UV lenticular filters, which cover the wavelength range between 1600~\AA\ and 6000~\AA, in addition to a clear white filter that covers the wavelength range between 1600~\AA\ and 8000~\AA\ \citep{pbp+08}. The data available to download at the \swift\ data archive\footnotemark[1] are reduced by the science data centre at Goddard Space Flight Center, and photometric analysis can be carried out immediately on the level 2 products, which are already in sky co-ordinates and aspect corrected. In order to convert UVOT images into spectral files compatible with the spectral fitting package, {\sc xspec}, we used the tool {\sc uvot2pha} (v1.1). The response matrices used for the UVOT filters were taken from the \swift /UVOT calibration files {\it swu**20041120v104.rsp}, where {\it **} is the code for the appropriate filter.

When ground-based optical or NIR data were available to use in the SED, spectral files were produced for each filter using the appropriate responsivity curves. Cousins $R$ and $I$ responsivity curves were taken from \citet{bes90}, and the $J$, $H$ and $K$-band responsivity curves were taken from \citet{cwb+92,cww92} and \citet{bcp98}. For the Sloan Digital Sky Survey (SDSS) $ugriz$ filters \citep{fig+96}, responsivity curves provided in the SDSS data release 6 were used\footnotemark[2].
\footnotetext[2]{http://www.sdss.org/dr6/instruments/imager/}

To produce an SED at an instantaneous epoch the magnitude of the afterglow at the epoch of the SED was measured by interpolating or extrapolating the UVOT and ground-based filter dependent light curves to the epoch of interest. The spectral files were then set to the extrapolated/interpolated magnitude measured in each corresponding filter.

For the UVOT filter light curves, source photometric measurements were extracted from the UVOT imaging data using the tool {\sc uvotmaghist} (v1.0) with a circular source extraction region that ranged from $3-5\arcsec$ radius, depending on the brightness of the source. In order to remain compatible with the effective area calibrations, which are based on $5\arcsec$ aperture photometry \citep{pbp+08}, an aperture correction was applied where necessary. The background was taken from a source-free region close to the target with a radius of between $10\arcsec$ and $20\arcsec$. The light curves were then binned into groups $\Delta T_{bin}/T = 0.1$, where $\Delta T_{bin}$ is the time interval of the bin, and $T$ is the time since the BAT trigger.

For the ground-based optical and NIR data, filter dependent light curves were produced using the data from the literature. Both for data taken from refereed publications, or from GCNs, which are subject to systematic uncertainties in absolute calibration, the calibration systematic error was added in quadrature to the photometric error on each measurement. Where no error was provided, either on the magnitude or calibration, an error of 0.3 magnitudes was assumed.

When interpolating or extrapolating each filter dependent light curve to the SED epoch, the same decay index was  fit over the same time interval for all the filter light curves within each GRB (for both UVOT and ground-based data). Both the time interval and decay index used were determined from the combined UVOT and ground-based light curve, where all filters were normalised to the UVOT white band, if available, and if not, to the v-band, to produce a single light curve. The time interval was chosen such that it covered the epoch of the SED and could be well-fitted by a power-law, and this ranged from $\Delta T/T_{SED}=0.5$ to $\Delta T/T_{SED}=5$, where $\Delta T$ is the time interval used, and $T_{SED}$ is the epoch of the SED. The best-fit decay index to this time interval was then used to fit each independent filter light curve. Having set the spectral files to the extrapolated/interpolated magnitude measured in each corresponding UVOT and ground-based filter, a further 10\% systematic error was added to each ground-based spectral data point to account for uncertainties in the responsivity curves.

\begin{figure*}
\centering
\includegraphics[width=1.0\textwidth]{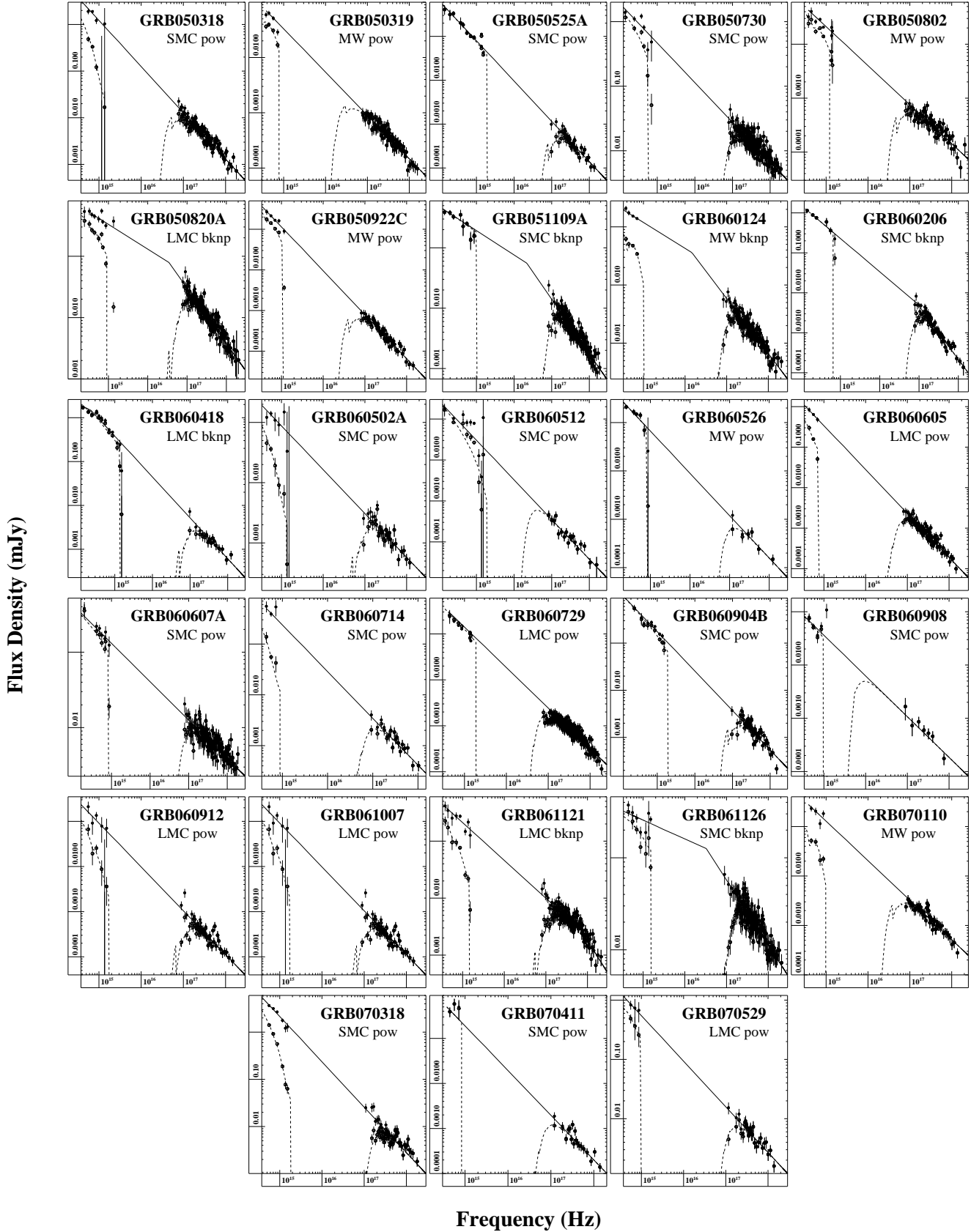}
\caption{The unabsorbed spectral energy distributions for the 28 GRBs in our sample in units of mJy against Hz. In each figure, we plot the best-fit absorption and extinction corrected spectral model (solid lines) and data (black data points), as well as the host galaxy absorbed and extinguished spectral model (dashed lines) and data (open data points). The extinction curve used in the fit is labelled for each SED, as well as the underlying continuum fit to the SED; either a power-law (pow) or a broken power-law (bknp).}\label{fig:SEDs}
\end{figure*}

\subsection{X-Ray Data}\label{ssec:xray}
The XRT is well-calibrated in the 0.3--10~keV energy range, and has two primary observing modes: window timing (WT), which has a 1.7~ms time resolution and 1-dimensional imaging, and photon counting (PC), which has a 2.5~s time resolution and full imaging capabilities \citep{bhn+05}. Both modes have spectroscopic capabilities. All data were reduced with the {\sc xrtpipeline} tool (v0.11.6) using the most current XRT calibration files, version 20080509. In most cases PC mode data were used, with the exceptions being GRB~060206 and GRB~061007, which only had WT mode data at the epoch of the SEDs (10~ks and 600~s after the BAT trigger, respectively). For PC data, both for the spectral and temporal analysis, source counts were extracted from a circular region centred on the source with a radius ranging from 20 to 64~pixels, where 1 XRT pixel is $2.36\arcsec$. When the source was piled-up we fitted the source PSF profile with XRT's known PSF \citep{mcm+05} to determine the radius at which pile-up becomes important, and used an annular extraction region to exclude data within this radius, which ranged from 4 to 6~pixels. The background count rate was estimated from a circular, source-free area in the field of view (FOV) with a radius ranging from 42 to 64~pixels. For WT mode data, the extraction regions used for the source and background were slits positioned over the source and in a source free region of the FOV, with lengths ranging from 20 to 80~pixels, respectively. For both PC and WT data, we used {\sc xselect} (v2.4\footnotemark[3]) to extract light curves and spectral files from the event data in the energy ranges 0.3--10~keV, which is the band required for compatibility with the current calibration files \footnotemark[4]. 
\footnotetext[3]{http://heasarc.nasa.gov/docs/software/lheasoft/ftools/xselect/}
\footnotetext[4]{http://heasarc.gsfc.nasa.gov/docs/heasarc/caldb/swift/docs/xrt/}
The spectral files were grouped to $\geq 20$ counts per energy channel, and the light curves were binned into time intervals of $\Delta T/T = 0.1$. Effective area files corresponding to the spectral files were created using the {\sc xrtmkarf} tool (v0.5.6), where exposure maps were taken into account in order to correct for bad columns. Response matrices from version 10 of the XRT calibration files were used for both WT and PC mode data. The spectra were normalised to correspond to the 0.3--10~keV flux of the GRB afterglow at the epoch of the SED. This flux was determined from the best-fit power-law decay model to the afterglow light curve, in the same way as was done for the UVOT and ground-based data. 

\section{The Model}\label{sec:model}
The SEDs were fitted within {\sc xspec} (v12.4.0\footnotemark[5]) using the same spectral models as those used in \citet{smp+07}, with the exception that in this paper absorption due to the Lyman forest is also accounted for.

For each SED we tried both a power-law and a broken power-law to fit the afterglow spectral continuum. In the latter case the change in spectral slope was fixed to $\Delta\beta = 0.5$ to correspond to the change in slope caused by the cooling frequency \citep{spn98} lying within the observed frequency range at the epoch of the SED. In both the power-law and broken power-law models, two independent dust and gas components were included to correspond to the Galactic and the host galaxy photoelectric absorption and dust extinction. The Galactic components were frozen to the column density and reddening values taken from \citet{kbh+05} and \citet{sfd98}, respectively, which although uncertain, in particular for lines-of-sight with large Galactic reddening, we found to be typically an order of magnitude smaller than the errors on the measured host galaxy absorption and extinction values. Furthermore, the uncertainty in the Galactic values becomes negligible when propagated to the additional error on the measured rest frame absorption and extinction values. The dependence of dust extinction on wavelength in the GRB host galaxy was modelled on the SMC, the LMC and the Milky Way (MW) empirical extinction laws using the {\sc xspec} tool {\sc zdust}, which is based on the extinction coefficients and extinction laws from \citet{pei92}. The total-to-selective extinction was taken to be \Rv\ = 2.93, 3.16 and 3.08 for the SMC, LMC and Galactic extinction laws, respectively \citep{pei92}. From here onwards we shall refer to each of the spectral models as the SMC, LMC and MW model, where the name corresponds to the extinction law used to describe the dust extinction properties in the GRB host galaxy. The equivalent neutral hydrogen column density in the host galaxy was determined from the soft X-ray absorption, where solar abundances were assumed, and is denoted throughout this paper as \nhx.
\footnotetext[5]{http://heasarc.nasa.gov/xanadu/xspec/manual/}

There have been a number of examples where intervening systems have been detected in the line-of-sight to GRBs (e.g. GRB~050730; \citet{sve+05,dfm+07}, GRB~050922C; \citet{pwf+08}, GRB~060418; \citet{evl+06,vls+07}, GRB~070802; \citet{efh+09}), although in the majority of cases, the dominant absorption system has been reported as originating at the host galaxy. The largest reported absorption from an intervening system to date corresponds to an absorption system at $z=2.077$ in the line-of-sight to GRB~050922C, which had a column density of  $N_{HI} = 2.0\times 10^{20}$~\invsqrcm \citep{pwf+08}, which is an order of magnitude less than the host galaxy neutral hydrogen column density. A notable exception to this was in the case of GRB~060418, which had a strong foreground absorber at $z=1.118$ with an estimated lower limit on the hydrogen column density of $N_{HI} > 1.7\times 10^{21}$~\invsqrcm\ and a marginally larger extinction than at the host galaxy. Nevertheless, the percentage of GRBs with reported strong intervening systems is small (e.g. Prochter et al. 2006, ApJ, 648, 93; Sudilovsky et al. 2007), and not including their intervening systems in our SED modelling is, therefore, unlikely to affect the overall results of this paper. 

To model the Lyman-series absorption in the 912--1215~\AA\ rest frame wavelength range, we wrote a local model for {\sc xspec}, and this was included in our fit of the afterglow SEDs. The model used the prescription provided in \citet{mad95} to estimate the effective optical depth from the Lyman-series as a function of wavelength and redshift. As well as estimating the hydrogen absorption caused by intervening systems, \citet{mad95} also determined the error on this due to statistical fluctuations in the number of absorbing clouds along the line-of-sight. This error was added in quadrature to the photometric uncertainty of any optical data at rest frame wavelengths blueward of Ly$\alpha$.

\section{Results}\label{sec:results}
\begin{figure}
\centering
\includegraphics[width=0.5\textwidth]{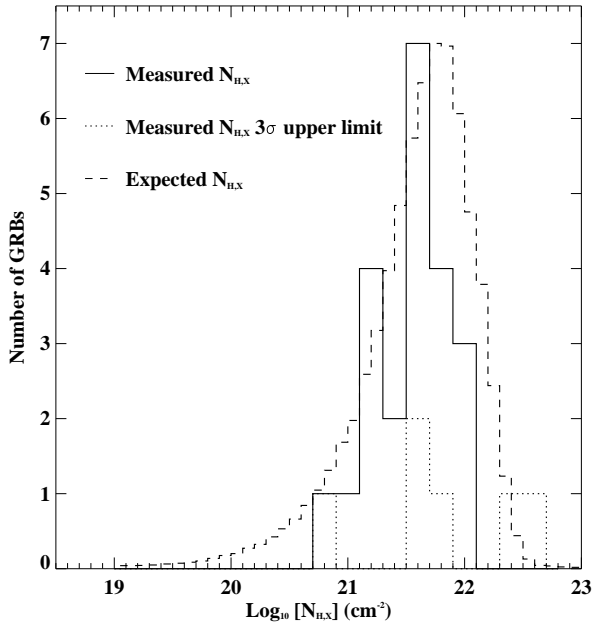}
\caption{Distribution of GRB host galaxy \nhx\ for those GRBs in the sample with a soft X-ray absorption system detected at the host galaxy at the $90$\% confidence level (solid histogram). For those GRBs with no detected host galaxy absorption system (90\% confidence), the distribution of \nhx\ $3\sigma$ upper limits is plotted (dotted histogram). The dashed histogram is the expected neutral hydrogen column density distribution for GRBs that occur within Galactic-like molecular clouds taken from \citet{rp02}.
}\label{fig:NHdist}
\end{figure}

\begin{figure}
\centering
\includegraphics[width=0.5\textwidth]{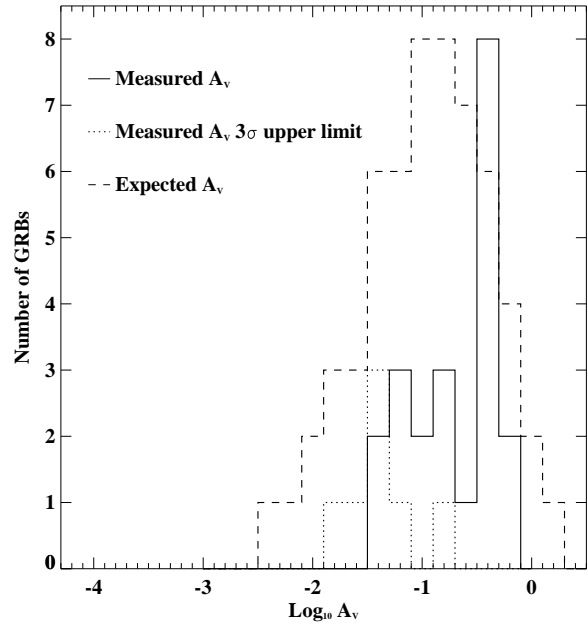}
\caption{Distribution of GRB host galaxy visual extinction for those GRBs in the sample with a dust extinction system detected at the host galaxy at the $90$\% confidence level (solid histogram). For those GRBs with no detected host galaxy extinction system (90\% confidence), the distribution of av\ $3\sigma$ upper limits is plotted (dotted histogram). The dashed line corresponds to the expected GRB host galaxy \av\ distribution when selection effects are taken into account (see section~\ref{ssec:seleffects}).
}\label{fig:AVdist}
\end{figure}

\subsection{Distribution of Host \av\ and \nhx}
\label{ssec:dist}
The results from our spectral analysis are summarised in Table~4, and the absorption corrected SEDs are shown in Fig.~\ref{fig:SEDs} in units of mJy against Hz. Each SED in Fig~\ref{fig:SEDs} shows the host galaxy absorbed and extinguished spectral model and data (dashed lines and open data points, respectively), as well as the best-fit absorption and extinction corrected spectral model and data (solid lines and black data points, respectively). We find that 79\% of the sample has a host galaxy absorption system detected with $90$\% confidence, with an equivalent neutral hydrogen column density, \nhx, ranging from $N_{H,X}=8.2\times 10^{20}$~\invsqrcm\ to $N_{H,X}=1.4\times 10^{22}$~\invsqrcm, and a dust extinction system local to the GRB is detected in 71\% of the sample with $90$\% confidence, with a range in extinction of $0.03<A_V<0.75$. We note here that the best-fit visual extinction is dependent on the shape of the extinction law used to fit the data, and \citet{ccm89} found a linear correlation between the total-to-selective extinction, \Rv, and the amount of UV extinction along different lines-of-sight within the Milky Way. For those GRBs best-fit by a broken power-law continuum, the additional free parameter in the fit introduces some degeneracy between the location of the spectral break and the total-to-selective extinction, \Rv, and in these cases, extinction laws with larger total-to-selective extinction, \Rv, than those considered in our analysis could result in larger best-fit extinction values than those listed in Table~4 \citep[e.g. see][]{wfl+06,efh+09}. If we, therefore, only consider those results from the GRBs best-fit by a power-law continuum, we find that the largest host galaxy visual extinction measured in the sample is still \av\ = 0.75.

Our measured distribution of host galaxy logarithmic column densities for GRBs with a host galaxy x-ray absorption system detected at 90\% confidence is shown in Fig.~\ref{fig:NHdist} in units of \invsqrcm (solid histogram), and has a mean of 21.7 and a standard deviation of 0.3. For those GRBs that do not have a host absorption system detected at 90\% confidence, the \nhx\ $3\sigma$ upper limit distribution is shown by the dotted histogram. As well as plotting the measured \nhx\ distribution, we also show the expected \nh\ column density distribution in the line-of-sight to GRBs within Galactic-like molecular clouds \citep[dashed histogram;][]{rp02}. The expected \nh\ distribution will only equate to an \nhx\ distribution in the case of solar metallicity GRB host galaxies. However, the host galaxies of GRBs typically have sub-solar metallicities that range from 1/100$^{th}$ solar to solar \citep{pcd+07,sgl09}. Combining the metallicity estimates from \citet{pcd+07} and \citet{sgl09} gives a distribution with median value of 1/4$^{th}$ solar, covering over two orders of magnitude. The effect of this on the expected GRB host galaxy \nhx\ distribution shown in Fig.~\ref{fig:NHdist} would be to broaden it to the left of the figure, down to values of $\log$~\nhx = 19, and shift the logarithm of the expected peak \nhx\ value down to 21.2. However, due to the small number of GRB host galaxies with accurate metallicity measurments, the GRB host metallicity distribution is poorly known. We therefore feel that for the time being it is not feasible to apply a correction to the expected \nh\ distribution shown in Fig.~\ref{fig:NHdist} to convert it into an accurate expected \nhx\ distribution.

Despite the above caveat with the expected \nhx\ distribution, it appears from Fig.~\ref{fig:NHdist} that those GRBs with host galaxy column densities at the very low and very high end of the distribution are missing from our sample. The lack of GRBs in our sample with measurements of host galaxy \nhx\ smaller than $\sim 10^{21}$~\invsqrcm\ is likely to be due to the sensitivity limit of the XRT at measuring soft X-ray absorption. In which case those GRBs with expected host galaxy \nhx\ smaller than $\sim 10^{21}$~\invsqrcm\ may be accounted for by those GRBs in our sample with just upper limits on their host galaxy soft x-ray absorption. The absence of GRBs in our sample with host galaxy $N_{H.X}\gtrsim 10^{22}$~\invsqrcm, on the other hand, is likely to be a consequence of our selection effects. By selecting only those GRBs with observed v $< 19$ we exclude highly extinguished GRBs from our sample, which would simultaneously exclude those with large X-ray absorption values. We shall address the effect that this bias has on our results later, in section~\ref{sec:seleffs}. Nevertheless, there is generally a good agreement between the two histograms, suggesting that our selection effects on the distribution of host galaxy column densities in the line-of-sight to GRBs is not highly significant.

The measured optical extinction distribution for GRBs with a host dust extinction system detected at 90\% confidence is shown in Fig.~\ref{fig:AVdist} (solid histogram), and has a mean \av\ of 0.3 with a standard deviation of 0.2. The dotted histogram shows the \av\ $3\sigma$ upper limit distribution for those GRBs with no host extinction system detected at 90\% confidence. In this figure we also show the expected GRB host galaxy \av\ distribution when selection effects are taken into account (dashed histogram), the details of which we describe in section~\ref{ssec:seleffects}.

\subsection{Host Galaxy Extinction Curves}
\label{ssec:extcurvetype}
The extinction properties of dust are dependent on the dust composition and grain size distribution. The  extinction curve models that best-fit the data, therefore, provide information on the dust properties of the GRB circumburst medium. For 18\% of the sample no distinction could be made in the quality of the fits between extinction curve models, and this is a consequence of the lack of absorbing dust in the local environments of these GRBs, as well as the low signal to noise of some of the data. Evidence of this is provided in Fig.~\ref{fig:AVNHX}, where we plot the GRB host galaxy \nhx\ against \av in $\log-\log$ space, using the best-fit values from the SMC (top panel), the LMC (middle panel), and the MW spectral model (bottom panel). Those GRBs where no distinction was possible between dust model fits are shown as grey circles, all of which only have upper limit measurements on the host galaxy extinction value.

For those GRB SEDs best-fit by a broken power-law, a degeneracy exists between the total-to-selective extinction, \Rv, and the location of the spectral break, thus limiting our knowledge of the shape of the host galaxy extinction law in these cases. Where the optical and X-ray afterglow emission lie on the same power-law component, however, the fewer number of variable parameters provides a greater handle on the shape of the host galaxy extinction law. Of those GRBs for which a distinction between model fits was possible, 21 were best-fit by a power-law spectral model, the large majority of which have a rest-frame wavelength range that safely covers the location of the 2175\AA\ absorption feature, and all have data blueward of 1500\AA\ in the rest-frame. The data available for this subset of GRBs therefore covers the wavelength range over which the three extinction laws modelled in this paper can be most effectively discriminated between, and we therefore use this subset of GRBs to study the distribution in best-fit extinction laws.

We find that the SMC extinction curve provided the best-fit in 56\% of cases, and the LMC and MW extinction curve both provided the best-fit to 22\% of cases. Of the four GRBs where the MW extinction law provided the best fit to the SED, three have optical data that span the wavelength range of the 2175\AA\ absorption bump at the rest frame of the host galaxy (GRB~050802, GRB~050922C, GRB~070110), making the detection of the 2175~\AA\ feature possible. Nevertheless, in all three cases the SMC and LMC spectral models also provide acceptable fits, and we, therefore, cannot claim a robust detection of the 2175~\AA\ absorption feature in the host galaxy of these three GRBs. The question of how typical the Milky Way absorption feature is in GRB host galaxies is beyond the scope of this paper, and an issue that we are looking to investigate in future work. That the host galaxy dust extinction properties for the majority of our sample are best-fit by the SMC extinction law is consistent with several previous studies in this field \citep[e.g.][]{sfa+04,kkz06,sww+07}. It is not yet clear what is responsible for the Milky Way absorption feature at 2175~\AA, although small carbonacious dust grains are thought to play an important role \citep [e.g.][]{dl84}, suggesting that such grains are not typical in the environments of GRBs, at least not once the GRB has occurred.

 \subsection{\nhx/\av\ Ratio In GRB Host Galaxies}
\label{ssec:gas2dust}
\begin{figure}
\centering
\includegraphics[width=0.5\textwidth]{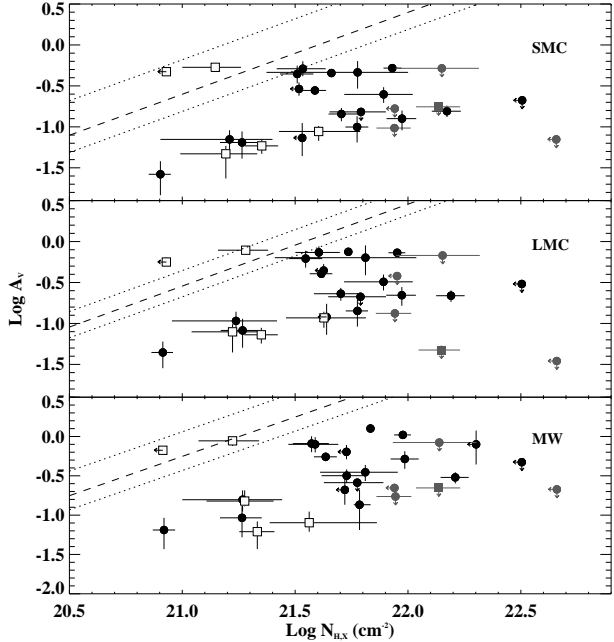}
\caption{Host galaxy \av\  against \nhx\ in $\log-\log$ space, taken from the spectral results from the SMC model (top panel), the LMC model (middle panel) and the MW model (bottom panel) spectral fits. The dashed and dotted curves are the \nhx/\av\ ratios and 1$\sigma$ deviations, respectively, for each corresponding environment, where we converted $N_H$ to an X-ray equivalent \nhx\ value assuming a metallicity 0.25 and 0.5 solar for the SMC and LMC, respectively (see text for details). The open squares correspond to those GRBs that were also used in the \citet{smp+07} sample, and the grey filled circles correspond to those GRBs for which no distinction can be made between the goodness of fit of the three spectral dust models.}\label{fig:AVNHX}
\end{figure}

In Fig.~\ref{fig:AVNHX}, the dashed lines in each panel represent the mean hydrogen column density to extinction ratios in the SMC \citep[top; ][]{mml89}, LMC \citep[middle; ][]{koo82,fit85} and Milky Way \citep[bottom;][]{ps95}, which have been converted from \nh/\av\ to an \nhx/\av\ ratio relating to the column density that would be measured from X-ray observations of each of these galaxies if solar abundances were assumed. We did this by assuming a metallicity of 0.25 solar for the SMC, and 0.5 solar for the LMC \citep{whm98}, and the \nhx/\av\ ratio was then a fraction 0.25 and 0.5 the \nh/\av\ ratio for the SMC and LMC, respectively. As well as the mean, we also show the root-mean square deviation (dotted lines) of the sample used to derive the mean SMC, LMC and Milky Way \nh/\av\ ratios, also converted into the equivalent \nhx/\av\ ratio. The subsample of GRBs that were analysed in \citet{smp+07} are shown as open squares.

From Fig.~\ref{fig:AVNHX} there does not appear to be any strong correlation between the dust and gas column density in GRB host galaxies. Using only those GRBs with an extinction and absorption system detected with $90$\% confidence, a spearman rank test between the best-fit \av\ and \nhx\ measurements from the SMC, the LMC and the MW models indicates a weak correlation at the $1\sigma$ level, with coefficients 0.39, 0.37 and 0.48, respectively. It is also notable that most of the data points lie to the right of the dashed lines, corresponding to \nhx/\av\ ratios that are larger than those of the SMC, LMC and MW. In \citet{smp+07} the GRB \nhx/\av\ ratios were also typically larger than those of the MW and Magellanic Clouds, although they were still consistent at the 68\% confidence level with the SMC, LMC and MW \nhx/\av\ ratios.

\begin{figure}
\centering
\includegraphics[width=0.5\textwidth]{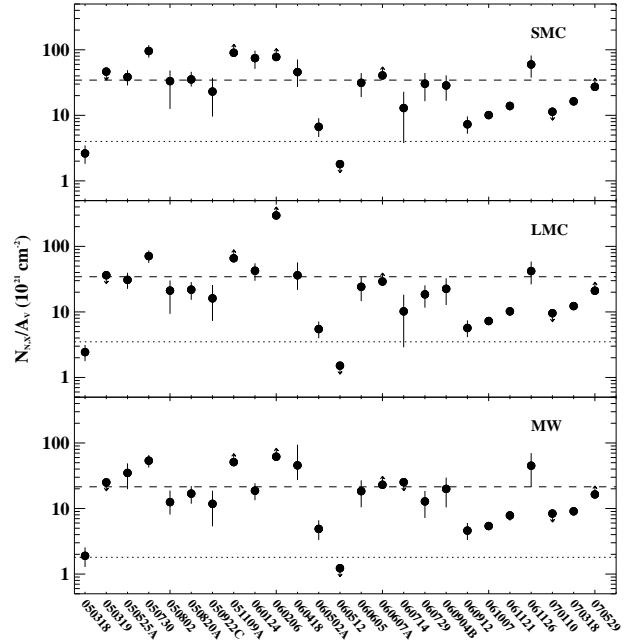}
\caption{Ratio of rest frame \nhx\ to \av\ resulting from the SMC (top panel), LMC (middle panel) and MW (bottom panel) extinction spectral models, where the results from the best-fit model between either the power-law or broken power-law fit are shown. The dotted lines correspond to the SMC, LMC and MW empirical \nhx/\av\ ratios, and the dashed lines indicate the mean \nhx/\av\ value from each set of extinction models fitted.}\label{fig:AVNHXrat}
\end{figure}

The larger sample used in this paper shows a spread in gas-to-dust ratios that covers nearly two orders of magnitude. This is better illustrated in Fig.~\ref{fig:AVNHXrat}, where we plot the \nhx/\av\ ratio from the SMC (top), LMC (middle) and MW (bottom) model fits for each GRB in the sample with a host galaxy absorption and/or extinction system detected with  $90$\% confidence. In each case the data points shown are the results from the best-fit model to the continuum. \ie\ either a power-law or broken power-law fit to the SED. The dashed lines are the mean \nhx/\av\ ratio of our sample determined from our spectral analysis for each of the extinction curve models fitted. These correspond to $\langle N_{H,X}/A_V\rangle = 3.3\times 10^{22}$~\invsqrcm, $3.4\times 10^{22}$~\invsqrcm\ and $2.1\times 10^{22}$~\invsqrcm, which is a factor of 8.3, 9.7 and 11.7 larger than the mean \nhx/\av\ ratios measured in the SMC, LMC and MW (dotted lines) in the top, middle and bottom panels in Fig.~\ref{fig:AVNHXrat}, respectively. The standard deviation of the data about the dashed lines is $2.8\times 10^{22}$~\invsqrcm, $5.9\times 10^{22}$~\invsqrcm\ and $1.8\times 10^{22}$~\invsqrcm\ for the SMC, the LMC and the MW spectral model results, respectively.

\section{Selection and systematic effects}
\label{sec:seleffs}
\subsection{Selection effects in \av}
\label{ssec:seleffects}
Those GRBs located in very dusty regions are less likely to be detected at optical wavelengths than those GRBs with a small host galaxy extinction, and could, therefore, be missing in our sample, as was pointed out in section~\ref{ssec:dist}. A possible indication of this selection effect is the small number of GRBs that have \av$\gtrsim 1$ in all three panels of Fig.~\ref{fig:AVNHX}. To ascertain better the impact of these selection effects on our results we compared our measured distribution of host galaxy \av\ and \nhx\ with the distribution resulting from a Monte Carlo simulation of 1000 GRBs with host galaxy neutral hydrogen column densities, \nhx, and host visual extinctions, \av\ taken at random, where a Gaussian \nhx/\av\ distribution was assumed.

\begin{figure}
\centering
\includegraphics[width=0.5\textwidth]{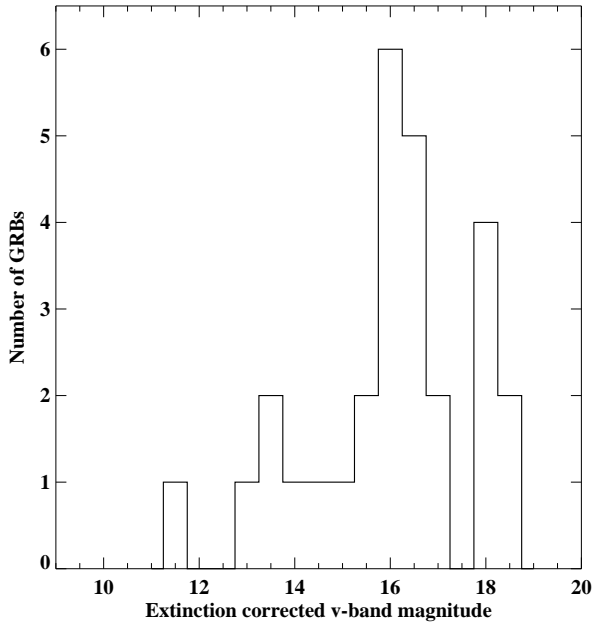}
\caption{Our sample distribution of the extinction corrected GRB v-band magnitude.}\label{fig:unextVdist}
\end{figure}

\begin{figure}
\centering
\includegraphics[width=0.5\textwidth]{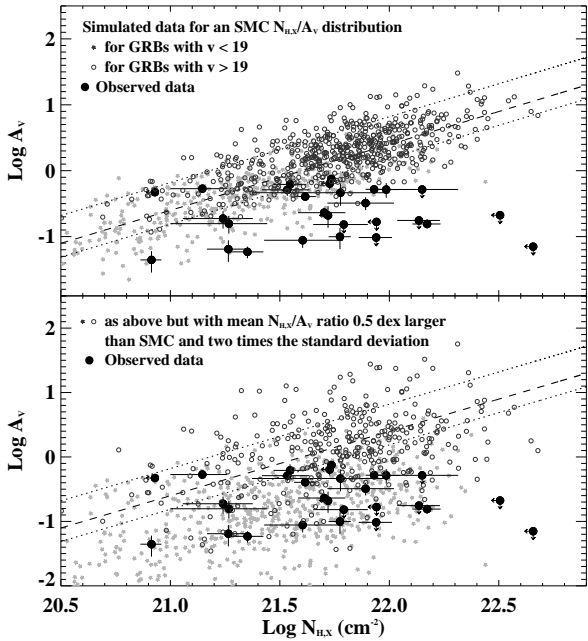}
\caption{Best-fit host galaxy \av\ against \nhx\ in log-log space for the sample of GRBs analysed in this paper (large circles), plotted together with a sample of randomly generated host galaxy \av\ and \nhx\ values. In the top panel a Gaussian \nhx/\av\ distribution with the same mean and standard deviation as observed in the SMC was assumed for the simulated data, and in the bottom panel a Gaussian distribution was again assumed, but this time with a mean \nhx/\av\ ratio 0.5~dex larger than that of the SMC, and two times the standard deviation. In both panels, those generated data with corresponding observed GRB v-band magnitudes v $< 19$ are plotted as stars, and those with v $> 19$ are shown as small open circles. The mean SMC \nhx/\av\ ratio (dashed line) and $1\sigma$ deviation (dotted line) are shown as a reference.}\label{fig:AVNHXvsMOC}
\end{figure}

We selected at random an equivalent neutral hydrogen column density, \nhx, from the expected GRB host galaxy \nhx\ distribution shown in Fig.~\ref{fig:NHdist}. For a given \nhx, the host galaxy visual extinction, \av, was then determined by assuming a random column density to visual extinction ratio, \nhx/\av, taken from a Gaussian distribution with a mean \nhx/\av\ ratio and standard deviation equal to that of the SMC. We chose to use the SMC \nhx/\av\ distribution since the majority of our sample are best-fit by an SMC host galaxy extinction law. In addition to selecting a host galaxy \nhx\ and \av, we also selected at random a redshift with $z\leq 4$ from the known \swift\ distribution, and a Galactic visual extinction and extinction corrected GRB apparent v-band magnitude from our sample distribution, shown in Fig.~\ref{fig:unextVdist}. With these parameters we could then determine the extinguished v-band magnitude that would be observed in each case, thus allowing us to calculate the fraction of simulated GRBs with observed v-band magnitudes v $> 19$ that would thus be rejected by our sample selection criteria. Using only those generated GRBs with observed v-band magnitudes v $< 19$, we then performed a two-dimensional Kolmogorov-Smirnov (KS) test between the simulated and measured \nhx\ and \av\ data sets, and found that they had less than a $1.2\times 10^{-5}$ probability of coming from the same parent population. The generated distribution of \av\ against \nhx\ is shown in log-log space in the top panel of Fig.~\ref{fig:AVNHXvsMOC} (small, open circles and triangles), together with our measured \av, \nhx\ values (solid circles). The poor agreement between the generated and observed data points can be clearly seen in this plot, with the generated data points typically lying above the observed data points.

We, therefore, re-ran our Monte Carlo simulation, but this time we used a Gaussian \nhx/\av\ distribution with a standard deviation two times that observed in the SMC, as well as trying a distribution with a mean \nhx/\av\ ratio 0.5~dex, 1.0~dex and 1.5~dex larger than that of the SMC, each with a standard deviation equal to and two times that observed in the SMC. The mean \nhx/\av\ ratios and standard deviations of the Gaussian distributions used in each of our Monte Carlo simulations are listed in Table~\ref{tab:gaussdis}, along with the results from a KS test between our measured \nhx\ and \av\ distribution and the simulated \nhx\ and \av\ data. In the last column of Table~\ref{tab:gaussdis} we also give the fraction of data points from each of the Monte Carlo simulations rejected by our selection criteria as a consequence of having an extinguished v-band magnitude v $> 19$.

\begin{table}
\begin{center}
\caption{Results from two-dimensional KS test between the observed host galaxy \nhx\ and \av\ distribution and the distribution simulated in eight separate Monte Carlo simulations. The simulated host galaxy \av\ values are calculated from a randomly selected host galaxy neutral hydrogen column density, \nhx, assuming an \nhx/\av\ ratio selected by random from a Gaussian distribution. The Gaussian \nhx/\av\ distribution assumed in each Monte Carlo simulation executed have different pairs of mean \nhx/\av\ ratio and \nhx/\av\ standard deviation, $\sigma$, given in columns 2 and 3 respectively. Columns 4 and 5 list the KS statistic, D, and KS probability, and in column 7 we give the fraction of GRBs rejected by our selection criteria due to having an observed magnitude $v > 19$.\label{tab:gaussdis}}
\begin{tabular}{@{}cccccc}
\hline
Model & mean & $\sigma$ & KS stat. & KS & Fraction of \\
number & \nhx/\av\ & & D & Prob & sample with\\
 & ($10^{22}$) & ($10^{22}$) & & & v $> 19$ \\
\hline\hline
1$^*$ & 0.4 & 1.0 & 0.622 & 0.000 & 57.20\% \\
2 & 0.4 & 2.0 & 0.430 & 0.007 & 57.30\% \\
3 & 1.3 & 1.0 & 0.311 & 0.095 & 31.10\% \\
4 & 1.3 & 2.0 & 0.306 & 0.108 & 35.20\% \\
5 & 4.0 & 1.0 & 0.344 & 0.047 & 85.80\% \\
6 & 4.0 & 2.0& 0.386 & 0.188 & 19.80\% \\
7 & 13 & 1.0 & 0.581 & 0.000 & 7.90\% \\
8 & 13 & 2.0 & 0.587 & 0.000 & 11.30\%\\
\hline
\end{tabular}
\footnotesize{$^*$  SMC mean \nhx/\av\ and standard deviation.}
\end{center}
\end{table}

Of all the \nhx/\av\ Gaussian distributions that we tried, we found that the two distributions of \nhx/\av\ with mean 0.5~dex larger than that of the SMC (models 3 and 4), and the two with mean \nhx/\av\ ratio 1.0~dex larger (models 5 and 6) all produced samples of \nhx\ and \av\ values that were consistant with our samples of host galaxy \nhx, \av\ measurements. In each of these cases the KS probability of the simulated and the measured data sets coming from the same parent population was at least 4\%. The simulated data sets that used an \nhx/\av\ distribution with the same mean \nhx/\av\ as that of the SMC (models 1 and 2), and a mean \nhx/\av\ ratio 1.5~dex larger than that of the SMC (models 7 and 8), had a KS probability of less than 1\% of being consistent with our observed sample. 

Another result of our Monte Carlo simulations is that they also provide the fraction of simulated GRBs rejected by our selection criteria. In a detailed analysis on 14 `dark\footnotemark[6]' 
\footnotetext[6]{Dark here refers to GRBs with no detected optical afterglow or with an optical flux that is significantly dimmer than expected from the observed X-ray afterglow.}
GRBs, \citet{pcb+09} estimated that at least 45\% of those GRBs with an apparent peak magnitude v $> 19$ were `dark' as a result of dust extinction. They also estimated that at least 20\% of all \swift\ GRBs have a host galaxy $A_V > 0.8$, and at least 10\% have $A_V > 2.5$. The host galaxy \av\ distribution produced by our model 4 is in closest agreement with the results from \citet{pcb+09}, generating a host galaxy visual extinction $A_V > 0.8$ 29\% of the time, and $A_V > 2.5$ 12\% of the time. Of all our Monte Carlo simulations, the fraction of GRBs rejected in model 4 by our selection criteria (35\%) was also the most consistent  with the fraction of GRBs with observed v $> 19$ estimated by \citet{pcb+09}. Given the similarity between the host galaxy \av\ and observed v-band distribution produced by our model 4 and the estimates from \citet{pcb+09}, we take the results from model 4 to be the most representative of the true host galaxy \nhx\ and \av\ distribution, and thus use these to quantify our selection effects.

In Fig.~\ref{fig:AVNHXvsMOC} we plot the distribution of simulated host galaxy \nhx\ and \av\ values resulting from model 4 in the bottom panel, together with model 1 in the top panel for comparison. Those GRBs with v $< 19$ are plotted as open stars, and those with v $> 19$ are shown as small open circles. Also shown are the best-fit host galaxy \nhx\ and \av\ values for our observed sample of GRBs (large circles), as well as the mean SMC \nhx/\av\ ratio (dashed line) and standard deviation (dotted line). The distribution of simulated data points for both GRBs with v $> 19$ and v $< 19$ shown in the bottom panel of Fig.~\ref{fig:AVNHXvsMOC} suggest that even when selection effects are taken into account, the majority of GRBs continue to have host galaxy \nhx/\av\ ratios that are larger than the SMC distribution. In fact, $\sim 80\%$ of simulated GRBs have \nhx/\av\ host galaxy ratios larger than the mean SMC value, and cover nearly four orders of magnitude in \nhx/\av. We therefore conclude that the results from our analysis that the distribution in the \nhx/\av\ ratio in GRB host galaxies is broad and typically larger than those of the Milky Way and Magellanic Clouds applies even when dust selection effects are taken into account. 

\subsection{Systematic effects in measuring \nhx}
\label{ssec:nhxsys}
Any spectral curvature present within the X-ray energy range, such as when the cooling frequency, $\nu_c$, or the prompt emission peak energy, $E_{pk}$, lies within the X-ray band, can result in an over-estimation of the measured \nhx\ if it is not taken into account in the spectral model fit. By fitting all our SEDs with a power-law continuum as well as with a spectral break corresponding to the cooling frequency, we tested for the possibility of $\nu_c$ lying within the observing band and used the results from the best-fit model. This, therefore, removes the probability that any spectral curvature resulting from $\nu_c$ lying within the X-ray band was incorrectly interpreted as soft X-ray absorption at the host galaxy. However, there may still be spectral curvature within the X-ray band if there was ongoing X-ray emission from the GRB during the time interval over which the spectrum was extracted. From the analysis of 59 GRBs, \citet{bk07} found such a prompt emission contribution out to a maximum of \T$+10^{4}$~s, after which none of the GRBs in their sample showed evidence of spectral evolution. The systematic effect that such spectral curvature would have on our SED analysis, therefore, only applies for those GRBs in our sample for which we produced SEDs at an epoch earlier than $T+10^4$~s. For the twelve GRBs in our sample for which this applies, we fit the hardness ratio for each GRB \citep{ebp+09} from the start of the time interval over which the XRT spectrum was extracted onwards, and found no evidence for spectral evolution over the time interval fitted. We therefore do not believe that spectral curvature in the X-ray band is systematically overestimating the host galaxy \nhx, in agreement with the results found by \citet{ngg+09}, who concluded that intrinsic curvature in the spectrum could not be considered as a general solution for the large GRB host galaxy \nhx.

\subsection{Systematic effects in measuring \av}
\label{ssec:avsys}
\subsubsection{Hydrogen absorption versus dust attenuation}
\begin{figure}
\centering
\includegraphics[width=0.5\textwidth]{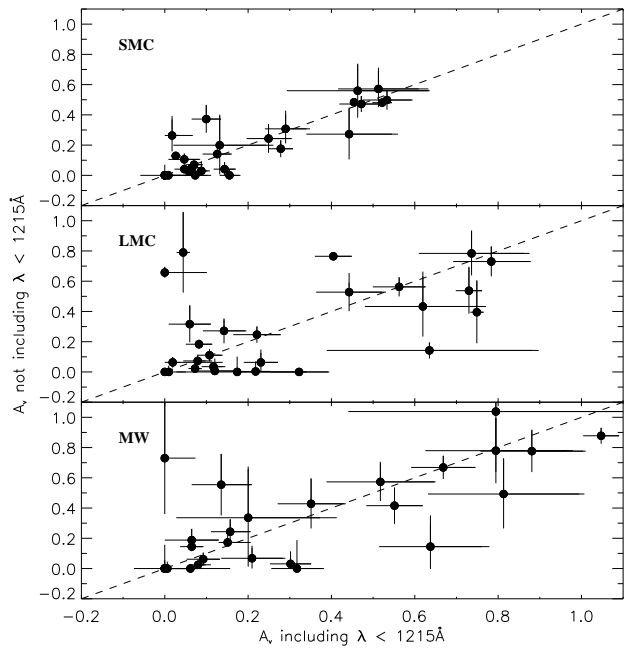}
\caption{Host galaxy \av\ resulting from spectral modelling of the SED where only those optical/NIR data redward of 1215~\AA\ in the rest frame were fitted, against the best-fit \av\ from fits to the complete SED. The spectral results for an SMC, LMC and MW host galaxy extinction law are plotted in the top, middle and bottom panels, respectively, and the dashed lines correspond to where there is no difference between the visual extinctions plotted along the $x$ and $y$-axes.}\label{fig:AVAV}
\end{figure}

If our Lyman forest model is overestimating the attenuation of UV light, this would result in a systematic underestimation of the extinction of UV light from dust. Therefore, to test the accuracy with which we model the Lyman-series absorption in our spectral analysis, we re-fitted the SEDs using only those optical/NIR data redward of 1215~\AA\ in the rest frame, where Lyman-series absorption no longer applies. In Fig.~\ref{fig:AVAV} we plot the best-fit dust-extinction yielded when optical data blueward of 1215~\AA\ were excluded, against the best-fit host galaxy dust-extinction values determined from the method outlined in section~\ref{sec:model}. The data points plotted in the three panels of Fig.~\ref{fig:AVAV} correspond to the best-fit dust-extinction values resulting from the SMC (top), LMC (middle) and MW (bottom) models, where the dashed lines correspond to where there is no difference between the visual extinctions plotted along the $x$ and $y$-axes. In all three panels the data points are evenly distributed about the dashed line. There is, therefore, no evidence to suggest that our modelling of the optical depth from the Lyman-series is resulting in a systematic effect on our best-fit \av\ values. The larger scatter around the dashed line in the bottom panel of Fig.~\ref{fig:AVAV} is due to the typically poorer fits given by the MW model compared to the SMC and LMC spectral models. It is worth pointing out that Fig.~\ref{fig:AVAV} also suggests that the lack of a host galaxy neutral hydrogen absorption component in our SED model is not overly affecting our results.

\subsubsection{Dust Extinction Cross-Section}
\label{ssec:crosssec}
By only considering the mean SMC, LMC and MW extinction laws in our spectral analysis, we may be introducing another systematic effect on our results. The amount of UV, optical and NIR radiation that is extinguished by dust is dependent on the density of dust in the environment \citep{pl02}, as well as the grain size distribution, the grain morphology, and the chemical composition, all of which influence the dust extinction cross-section as a function of wavelength \citep[e.g][]{pei92}. The measured visual extinction, $A_V$, will therefore depend strongly on the extinction law fit to the data \citep[e.g.][]{ccm89}. This is illustrated by the differences in the best-fit \av\ values measured between the spectral fits to the GRB SEDs. The total-to-selective extinction, \Rv, in the local environment of the GRB may be larger than the three mean values that we have considered \citep[e.g.][]{pbb+08}, representative of an environment with a dust size distribution skewed to larger grains. This would produce an extinction curve that is flatter in the NIR wavelength range than the SMC, LMC and MW extinction laws. Such an extinction law could account for the discrepancy that exists between the small amount of reddening observed in GRB SEDs \citep[e.g.][]{spl+05}, and the larger host galaxy dust column densities derived from dust depletion studies \citep[e.g.][]{sff03}. Such an extinction law is caused by a dust distribution skewed towards larger grains, which may result if there is ongoing dust destruction, or if the GRB itself preferentially destroys the smaller dust grains during its initial outburst. In the case where the dust in the GRB surrounding environment has such a grain size distribution, modelling the optical afterglow SED with an SMC extinction law would underestimate \av, since the steepness of the SMC extinction law over the NIR, optical and UV range, would yield a smaller value of \av\ for the same amount of UV extinction.

An estimate of the host galaxy visual extinction that does not require knowledge of the host galaxy extinction law is provided from GRB optical spectral analysis, where measured metal column densities and dust depletion models are used to estimate the fraction of metals locked up in grains. From the analysis of three GRB optical spectra, \citet{sff03} measured a mean host galaxy visual extinction of $\langle A_V\rangle \sim 1.0$, which is several times larger that the typical host galaxy extinction values measured from SEDs \citep[][]{sfa+04,sww+07,kkz06}, such as in this paper. One explanation for this difference in the visual extinction estimates between SED and optical spectroscopic analysis could be the presence of grey dust, which produces a flat extinction law. In such a case the near-uniform dust extinction across the optical and UV wavelength range would leave the observed optical/UV spectral slope relatively unchanged, and this may thus result in an underestimation of the best-fit \av\ when fitting the SED. However, in the case where the GRB SED is best-fit by a single power-law component, the reduced effect of dust-extinction on the X-ray and NIR bands allows the underlying spectral index to be well pinned, and in such a case grey extinction should be well detected. Of the 28 GRBs in our sample, 21 were statistically better-fit by a power-law spectral model, and the majority of our sample should therefore have well-determined measurements of the host galaxy extinction. Furthermore, the results from our MW extinction model, which has a flatter extinction law than the SMC and LMC, and hence yields a larger \av\ value for a given amount of dust absorption in the UV, still gives \nhx/\av\ ratios that are around an order of magnitude larger than that observed in the MW.

Another important point to consider is that extinction estimates from optical spectroscopic analysis require certain assumptions to be made on the local environment of the GRB, such as the dust depletion pattern and ionisation state of the surrounding gas. In \citet{sff03}, both the GRB host galaxy dust depletion chemistry and the \nh\ to \av\ ratio were assumed to be the same as in the Milky Way. In particular, if an LMC or SMC \nh/\av\ ratio were assumed, the \av\ estimates in \citet{sff03} would decrease by a factor of more than 3 and 8, respectively. \citet{pcd+07} estimated the GRB host galaxy extinction for a sample of GRBs using the metal column densities that they measured in the GRB optical spectra, but they assumed an SMC gas-to-dust ratio, and for those GRBs where a visual extinction estimate was possible, they estimated a maximum extinction value of \av = 0.18.

Finally, even if we were to adopt an extinction of $A_V\sim 1.0$, as estimated by \citet{sff03}, it is still considerably smaller than would be expected in environments with similar \nhx/\av\ ratios as in the Milky Way and Magellanic Clouds given the mean GRB host galaxy column density, \nhx. GRB host galaxy extinctions would need to be around an order of magnitude larger than the values that we measure in our analysis in order to be consistent with the \nhx/\av\ ratios measured in the Milky Way and Magellanic Clouds (see Fig.~\ref{fig:AVNHXrat}).

\section{Discussion}\label{sec:disc}
In \citet{smp+07} we proposed that the general consistency between GRB host galaxy gas-to-dust ratios and those of the Milky Way and Magellanic Clouds was evidence that the level of photo-ionisation caused by the GRB was of the same order as the amount of dust destroyed by the prompt emission, as indicated in \citet{pl02}. However, our analysis presented in this paper on a sample four times the size indicates that GRB host galaxies have typically larger gas-to-dust ratios than those of the SMC, LMC and MW. This is partly as a result of the larger sample in this paper, but is also due to the inclusion of the UV and red data in the SED, which has improved the determination of the rest frame extinction. In two cases our choices of SED epoch in this paper have also improved the analysis, where in \citet{smp+07} the epochs of our SEDs for GRB~050525A and GRB~050802 were at times when \swift\ data were poorly sampled. Nevertheless, the host galaxy \nhx\ and \av\ values determined in this paper are consistent at 90\% confidence with the best-fit host \nhx\ and \av\ values in \citet{smp+07}. 

In the rest of this section we shall explore the reasons that could account for the relatively large gas-to-dust ratios measured in GRB host galaxies. In section~\ref{ssec:dustdest} we investigate the effect that the GRB has on its surrounding environment, in section~\ref{ssec:nhxvsnh} we look at the differences in the regions of gas probed by X-ray and optical observations, and in section~\ref{ssec:dirrs} we explore the effect that the metallicity of GRB host galaxies has on the \nhx/\av\ ratios.

\subsection{Effect of GRB on Local Environment}
\label{ssec:dustdest}
To investigate how the GRB prompt emission alters its surrounding environment, and in particular how it affects X-ray and optical observations of the afterglow, Perna \& Lazzati (2002; here onwards PL02) simulated the photo-ionisation and dust destruction caused by a GRB within a molecular cloud. The more dust in the line-of-sight that is destroyed, and the more gas that is photo-ionised, the smaller the measured values of \av\ and \nhx, respectively. Differences in the efficiency of the dust destruction and photo-ionisation can bring about an overall change in the \nhx/\av\ ratio measured before and after the GRB event.

PL02 modelled the effect that a GRB would have on its surrounding environment when embedded within a Galactic-like molecular cloud with a column density of $N_H = 10^{22}$~\invsqrcm\ and \av = 4.5~mag, and varied the particle density by changing the radius of the cloud from $R = 10^{18}$~cm to $R = 10^{20}$~cm. They then simulated the soft X-ray absorption and visual extinction that would be measured with time along the line-of-sight to the GRB during its prompt emission phase, as the high energy radiation photo-ionised the gas and destroyed the dust within the molecular cloud. They found that in a large and diffuse region, photo-ionisation is more efficient than dust destruction, and as a result, the \nhx/\av\ ratio measured after the GRB would be smaller than the value prior to the GRB. On the other hand, as the density becomes larger and the region more compact, dust destruction gradually becomes more efficient with respect to the photo-ionisation, and the \nhx/\av\ ratio measured after the GRB is thus larger than the initial value. They repeated their work using a GRB with a softer spectrum, and found a similar result but with the cross-over, when dust destruction becomes more efficient than the photo-ionisation, occurring at a lower circumburst density. PL02 also investigated how an increase by a factor of 50 in the density within a cloud of radius $R = 10^{19}$~cm would effect the X-ray absorption and optical extinction, and they found that there was a greater difference in the efficiency in the dust destruction and photo-ionisation processes, where the former was the more effective.

The relatively large gas-to-dust ratios measured in our sample of GRBs could, therefore, be a consequence of them being embedded in dense molecular clouds with column densities on the order of $N_H = 10^{23}$~\invsqrcm, which has already been suggested by several authors \citep[e.g.][]{gw01,rp02,crc+06,vmz+04}. If dust destruction is the cause for the large \nhx/\av\ ratios, then the range observed in \av\ and \nhx\ values for GRB host galaxies could result from differences in the initial column densities and sizes of the molecular cloud. A further parameter to consider is the location of the GRB within the cloud, where GRBs located closer to the outer edges nearest to the observer would destroy and photo-ionise a greater fraction of dust and gas along the line-of-sight than a GRB embedded deeper within the molecular cloud. 

One consequence of dust destruction by the GRB is the colour evolution of the afterglow as the dust is destroyed. This is because smaller dust grains are destroyed more easily than larger ones, such that the opacity to blue light will decrease sooner than the opacity to red light. However, in order to observe the spectral evolution caused by dust destruction, multi-wavelength observations in the optical and infrared range are required during the dust destruction period. During the first $\sim 600$~s of the UVOT observing sequence followed when a GRB occurs, only two filters are used. The multi-wavelength observations that are necessary to observe any spectral evolution in the UV-NIR afterglow therefore only begin $\sim 600$~s after the BAT trigger, by which time the high energy emission responsible for the dust destruction is over \citep{fkr01}. Robotic ground-based telescopes, such as the Rapid Eye Mount \citep[REM;][]{zcg+04}, and the RAPid Telescope for Optical Response \citep[RAPTOR;][]{sm08}, can be taking multi-wavelength data of a GRB a mere  $\sim 20$~s after the GRB trigger \citep[e.g.][]{acd+08}. However, thus far there have not been any clear instances of GRB light curves with observed colour evolution resulting from dust destruction. It may be possible to detect evidence of dust destruction in single filter observations by the increase in the observed flux within that filter that should observed as the dust opacity decreases. However, in a detailed analysis on the optical early-rise of a sample of six GRBs, five of which are in the GRB sample studied in this paper, \citet[e.g.][]{ops+09} found no evidence of the brightening of the afterglow to be the result of dust destruction for any of the GRBs in their sample.

It is likely that even at 20~s after the prompt emission, the bulk of the dust destruction has already occurred (PL02), making such direct observations of dust destruction highly challenging for the current generation of telescopes and satellites. Therefore, another way of identifying a recent period of dust destruction is in the grain size distribution in the local environment of the GRB. A consequence of dust destruction is a grain size distribution skewed towards larger grains due to the preferential destruction of small dust grains, which will result in a flattening of the extinction law \citep{plf03}. Although large dust grains may be broken down into smaller grains, these would, subsequently, be shattered promptly by the GRB, keeping the population of small dust grains small \citep[e.g.][]{plf03}. The best-fit extinction laws to our sample of GRBs should therefore provide an indication of the grain size distribution of the extinguishing dust within the GRB environment. The fact that 56\% of GRBs with a well-constrained host galaxy extinction law are best-fit by the SMC extinction law (see section~\ref{ssec:extcurvetype}), which has the steepest UV extinction of all the extinction curves fitted to our data, indicates that there is an abundance of small dust grains in the GRBs' surrounding environments. The flattest extinction law that we fitted to our data was the MW extinciton law, which also has the most prominent 2175\AA\ absorption feature. It is, therefore, possible that the typically poorer fits provided by the MW model are due to the absence of the 2175\AA\ absorption feature in GRB SEDs, rather than a poor agreement between the slope of the MW extinction law and the GRB hosts' extinction law. However, the LMC extinction law is flatter than the SMC extinction law, and has a relatively weak 2175\AA\ feature, and yet it was the best-fit model to only 22\% of the sub-sample of GRBs with well-constrained power-law spectral fits. The fact that more than two times as many GRB SEDs are better fit with an SMC extinction law than with an LMC extinction law for the host galaxy suggests that there remains an abundance of small, UV absorbing dust grains in the surrounding environments of GRBs. The dust responsible for the UV and optical extinction must, therefore, lie in regions of the GRB host galaxy that have not been subjected to significant amounts of dust destruction.

\subsection{Comparison between \nh/\av\ and \nhx/\av}\label{ssec:nhdisc}
\label{ssec:nhxvsnh}
\nhx\ provides a measurement of the optical depth of metals in the line-of-sight to the GRB, and in comparing the neutral hydrogen column density, \nh, with \nhx\ for a sample of 17 GRBs, \citet{whf+07} found \nh\ to be systematically smaller than \nhx. They ascribed this to X-ray absorption and Ly$\alpha$ absorption observations probing different regions of gas. Estimates of the distance of the UV/optical absorbing dust  from GRBs range from 100~pc to 1.7~kpc \citep{pcb06,vls+07}, whereas there are likely to be partially ionised medium weight metals absorbing the soft X-ray emission within a few parsecs of the GRB \citep{fkr01}. Although this does not necessarily mean that all the neutral hydrogen out to 100~pc is ionised, it does suggest that all the neutral hydrogen within the molecular cloud surrounding the GRB has been ionised. So whereas \nh\ will probe the host galaxy ISM outside of the molecular cloud, \nhx\ will, in addition, probe the gas within the molecular cloud, where densities of partially ionised oxygen and other medium weight metals remain relatively high. These differences in the gas probed by \nh\ and \nhx\ measurements provide information on the conditions of the environment at varying radii from the GRB.  Having compared \av\ to \nhx, we now look at the relation between \av\ and \nh\ to investigate the relative location of the regions of dust and gas probed by these two measurements.

\begin{figure}
\centering
\includegraphics[width=0.5\textwidth]{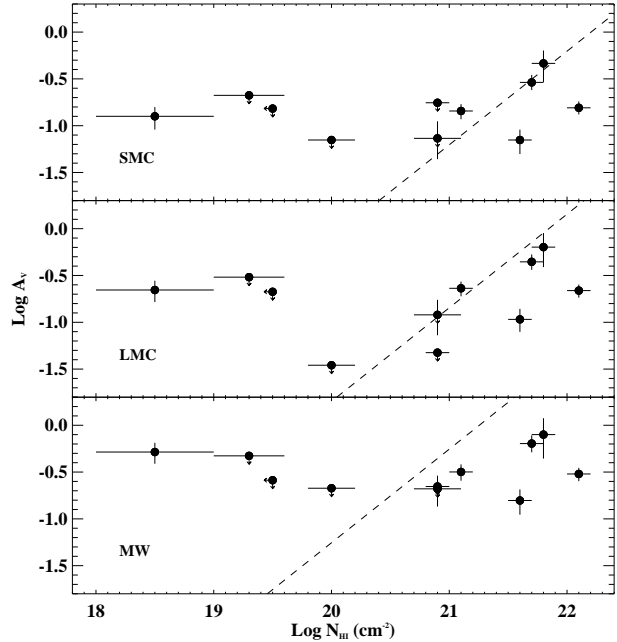}
\caption{Logarithmic host galaxy \av\ against logarithmic \nh\ for a subsample of GRBs with \nh\ measurements available in the literature. \nh\ values are taken from \citet{jfl+06a} and references therein. The \av\ values are the results from our spectral analysis from the SMC model (top panel), the LMC model (middle panel) and the MW model (bottom panel). The dashed curves are the \nh/\av\ ratios for each corresponding environment.}\label{fig:AVNH}
\end{figure}

We took \nh\ values reported in the literature for all GRBs that overlapped with our sample (Jakobsson et al. 2006 and references therein), of which there are eight (see Table~\ref{tab:nh}). Fig.~\ref{fig:AVNH} shows \av\ from the SMC (top panel), LMC (middle panel) and MW (bottom panel) spectral model fits plotted against \nh, in $\log-\log$ space. The dashed line in each panel, from top to bottom, represent the \nh/\av\ ratios for the SMC \citep{mml89}, LMC \citep{koo82,fit85} and MW \citep{ps95}, respectively. 

\begin{table}
\begin{center}
\caption{Subsample of GRBs with \nh\ and metallicity, $[M/H]$, measurements available in the literature.\label{tab:nh}}
\begin{tabular}{@{}lcc}
\hline
GRB & $\log$~\nh & [M/H]\\
\hline\hline
050319 & $20.9\pm 0.2^1$ & - \\
050730 & $22.1\pm 0.1^1$ & -2.26$\pm 0.14^3$ \\
050820A & $21.1\pm 0.1^1$ & -0.63$\pm 0.11^3$ \\
050922C & $21.6\pm 0.1^1$ & -2.03$\pm 0.14^3$ \\
060124 & $18.5\pm 0.5^2$ & - \\
060206 & $20.9\pm 0.1^1$ & -0.85$\pm 0.18^3$ \\
060418 & - & -1.65$\pm 1.00^3$\\
060526 & $20.0\pm 0.2^1$ & -1.09$\pm 0.24^4$ \\
060607A & $< 19.5^1$ & $-$ \\
060714 & $21.80\pm 0.10^2$ & $-$ \\
070110 & $21.70\pm 0.10^2$ & $-$ \\
070411 & $19.30\pm 0.30^2$ &  $-$ \\
\hline
\end{tabular}
\footnotesize{

$^1$  \citet{jfl+06a}\\
$^2$ \citet{fjp+09}\\
$^3$ \citet{pcd+07}\\
$^4$ \citet{tkj+08}
}
\end{center}
\end{table}

For each GRB plotted in Fig.~\ref{fig:AVNH}, \nh\ is typically an order of magnitude smaller than \nhx\, as was noted in \citet{whf+07}, and the GRB \nh/\av\ ratios span at least an order of magnitude to each side of the Milky Way and Magellanic Cloud \nh/\av\ ratios. This is in contrast to the GRB \nhx/\av\ ratios, which were mostly larger than those of the Milky Way and Magellanic Clouds. A spearman rank test between \nh\  and the best-fit \av\ from the SMC, the LMC and MW models gives the spearman coefficients 0.34, 0.25 and 0.09 respectively, indicating that there is only a weak correlation, if any, between these two parameters.

Where the GRB data points lie relative to the dashed lines in Fig.~\ref{fig:AVNH} is the result of two competing effects. Host galaxy metallicities that are smaller than the Magellanic Clouds, as well as any significant dust destruction caused by the GRB, will move the data points downwards, below the dashed lines, and counteracting this effect will be the amount of photo-ionisation of hydrogen in the surrounding environment of the GRB that will move the data towards the left of Fig.~\ref{fig:AVNH}. The mean logarithmic metallicity of those GRBs plotted in Fig~\ref{fig:AVNH} with known metallicity is $\langle [M/H]\rangle = -1.37$ (i.e. $0.04~Z_{\odot}$), which is almost 1.0~dex smaller than the SMC, in which case we would expect the majority of the data points in Fig.~\ref{fig:AVNH} to lie below the dashed lines. The fact that the data points are fairly evenly distributed about the dashed lines, therefore, indicates that a greater volume of gas has been photo-ionised by the GRB, thus reducing \nh, than the volume of dust destroyed, which reduces \av. This is consistent with the analysis of high-resolution spectroscopic data, which indicates that GRBs can photo-ionise all gas within the molecular cloud surrounding the GRB \citep{pcd+07,vls+07}, whereas a GRB is only expected to fully destroy dust out to a few parsecs at most \citep{pl02}. The typically larger values of \nhx\ compared to \nh\ discussed by \citet{whf+07}, and the distribution in \nh/\av\ plotted in Fig.~\ref{fig:AVNH}, therefore, suggest that measurements of \nhx\ and \av\ probe regions of gas and dust within the molecular cloud, much closer to the GRB than measurements of \nh.

\subsection{GRB host galaxies and Dwarf Irregulars}
\label{ssec:dirrs}
\begin{figure}
\centering
\includegraphics[width=0.5\textwidth]{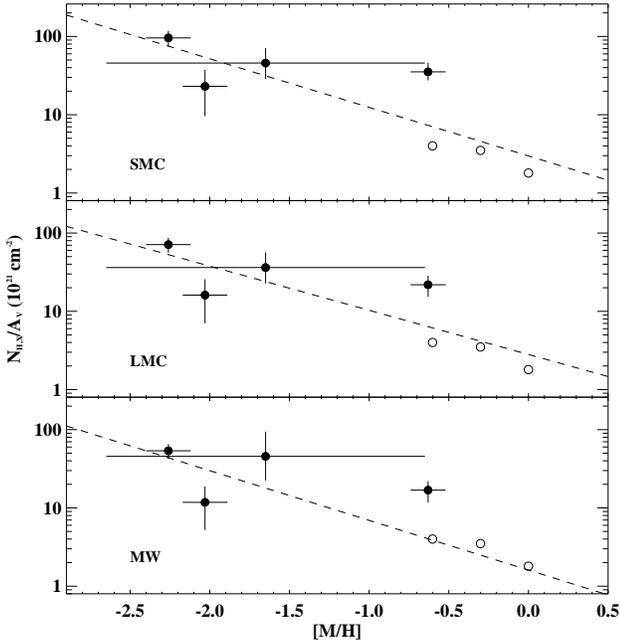}
\caption{Host galaxy \nhx/\av\ against metallicity, $[M/H]$, for a subsample of five GRBs (solid circles) with $[M/H]$ values available from the literature (see Table~\ref{tab:nh}) and a soft X-ray absorption system and/or dust extinction system detected with $90$\% confidence. The \nhx\ and \av\ values are the best-fit parameters from the SMC (top), LMC (middle), and MW (bottom) spectral fits. Open circles correspond to the SMC, LMC and Milky Way from left to right, respectively. The dashed line is the line of best-fit to the SMC, LMC, Milky Way, and the four GRBs with a soft X-ray absorption and dust extinction system detected with $90$\% confidence.
}\label{fig:NHXAVvsZ}
\end{figure}

It has been noted that for a large range of galaxy types, the gas-to-dust ratio of galaxies is inversely proportional to the metallicity, down to the most metal-poor systems \citep{ddb+07}. The range in column density to extinction ratio in GRB host galaxies may, therefore, be accounted for if those host galaxies with lower metallicities have larger \nhx/\av\ ratios. Evidence of such a correlation in our GRB sample is indicated in Fig.~\ref{fig:NHXAVvsZ}. Here we have plotted \nhx/\av\ against the host metallicity, $[M/H]$, for a subsample of GRBs (solid circles) that have an estimate of the host metallicity as well as a host galaxy soft X-ray absorption system and/or a dust extinction system detected with 90\% confidence. The top, middle and bottom panels show the \nhx/\av\ ratios determined from the SMC, LMC and MW spectral model fits, respectively. Those GRBs in our sample with host metallicity measurements are listed in Table~\ref{tab:nh}, along with their metallicity and corresponding reference. Although listed in Table~\ref{tab:nh}, GRB~060206 and GRB~060526 are not included in Fig.~\ref{fig:NHXAVvsZ}, since they only have upper limits for both \nhx\ and \av. The open circles in Fig.~\ref{fig:NHXAVvsZ} correspond to the SMC, LMC and Milky Way from left to right, respectively.

If very low metallicities are the over-riding reason that GRB host galaxies typically have larger \nhx/\av\ ratios than the Magellanic Clouds and Milky Way, then there should be a correlation between the metallicity and \nhx/\av\ ratio for the combined sample of GRB host galaxies and other, more metal-rich galaxies. A spearman rank test between the \nhx/\av\ ratio and the metallicity, [M/H], for the four GRBs shown in Fig.~\ref{fig:NHXAVvsZ}, together with the SMC, LMC and MW data points gives a coefficient of -0.89 with $90$\% confidence for each of the spectral models. This indicates a strong anti-correlation with a high level of significance. The dashed line in each panel is the best-fit power-law to the data.

It such a correlation is confirmed in the future with a greater sample of GRB host galaxies with a measured metallicity and \nhx/\av\ ratios, this would imply that low-metallicity galaxies are less efficient at forming dust from their metals than high-metallicity galaxies. One possible cause of this is an increase in supernovae dust destruction efficiencies in low metallicity environments, resulting from intermittent periods of star formation \citep{htk02}.

\section{Summary \& Conclusions}\label{sec:cons}
In this paper we have presented the results from the spectral analysis of 28 GRB SEDs. We measured the equivalent neutral hydrogen column density and visual extinction at the host galaxy, and found 79\% of the GRBs in our sample to have a detectable soft X-ray absorption system in the host galaxy, and 71\% to have a detectable visual dust-extinction system. Using the measured \nhx/\av\ ratios as an indicator of the host galaxy gas-to-dust ratio, we find that GRB host galaxies have gas-to-dust ratios that are typically larger than those measured in the Milky Way and Magellanic Clouds by up to two orders of magnitude. We have investigated several possibilities that could account for the relatively large gas-to-dust ratios in GRB host galaxies.

There is no evidence to suggest that the large host galaxy \nhx/\av\ ratios measured in our GRB sample is the result of any systematic error in the way that we measure \av. One possibility is that dust destruction by the GRB has reduced the visual extinction, \av, relative to the equivalent neutral hydrogen column density, \nhx. However, there are currently no observations that clearly show the early time colour evolution expected from dust destruction. Although such observations are limited by the quality and promptness of the data, we also found that the majority of our sample had host dust properties best-fit by the UV steep, SMC extinction law, indicating an abundance of small dust grains in the GRB surrounding environment. In the event of a significant phase of dust destruction, a grey extinction law should be observed, where the differential change in extinction from UV to NIR energy bands is small. The dust probed by our \av\ measurements must, therefore, lie in regions of the GRB host galaxy that have not been subjected to significant amounts of dust destruction.

For a subset of eight GRBs we were also able to study how the neutral hydrogen column density, \nh, compared with \av, and we found \nh/\av\ to extend to both larger and smaller values than those of the Magellanic Clouds and the Milky Way by up to an order of magnitude. The distribution in \nh/\av\ can be accounted for by the competing effects that alter the values of \nh\ and \av. Firstly, differences in the host galaxy metallicities and in the amount of dust destroyed by the GRB will affect the value of \av. On the other hand, the value of \nh\ will be dependent on the amount of photo-ionised hydrogen along the line-of-sight to the GRB. The mean logarithmic metallicity of the GRB sample with both \nh\ and \av\ measurements is almost 1.0~dex smaller than that of the SMC (0.04~$Z_{\odot}$), and we would therefore expect the GRB host galaxy \nh/\av\ ratio to be significantly smaller than the SMC \nh/\av\ ratio. The roughly even number of GRBs with smaller and larger \nh/\av\ ratios than the Magellanic Clouds and Milky Way therefore implies that the level of photo-ionised hydrogen along the line-of-sight to the GRB is greater than the fraction of dust destroyed by the GRB. This would suggest that measurements of \nhx\ and \av\ probe regions of dust and gas much closer to the GRB than \nh.

It has been suggested that differences in the gas-to-dust ratios in galaxies of different types are correlated with the metallicity of the galaxy \citep[e.g.][]{ddb+07}, whereby smaller metallicity systems have larger gas-to-dust ratios. From a subsample of four GRBs with measured metallicity and a soft X-ray absorption and visual extinction system detected with $90$\% confidence, together with the Small and Large Magellanic Clouds and Milky Way, we found a strong negative correlation between the \nhx/\av\ ratio and the metallicity, [M/H]. The spearman rank coefficient was -0.89 with $90$\% confidence. The large \nhx/\av\ ratios measured in GRB host galaxies could, therefore, be an indication of their very low, although broad, range of metallicities. A greater sample of GRB hosts with measured metallicities are needed to verify such a correlation, which if confirmed would suggest that low-metallicity environments are less efficient at forming dust from their metals than high-metallicity galaxies.

\section*{ACKNOWLEDGEMENTS}
We thank the anonymous referee for very helpful comments that have improved the quality of this paper, and we also gratefully acknowledge the contribution of all members of the \swift\ team. This research has made use of data obtained from the High Energy Astrophysics Science Archive Research Center (HEASARC), the UK Swift Science Data Centre at the University of Leicester, and the Leicester Data base and Archive Service (LEDAS), provided by NASAÕs Goddard Space Flight Center and the Department of Physics and Astronomy, Leicester University, UK, respectively.

\onecolumn

\appendix
\begin{table*}
\centering
\begin{minipage}{126mm}
\setcounter{table}{3}
\caption{Results from simultaneous UV/optical and X-ray spectral fits for the SMC, LMC and MW dust-extinction law models, for both a power-law (pow) and a broken power-law (bknp) continuum. The third and fourth columns give the host galaxy equivalent column density and visual extinction, the fifth column gives the break energy for the broken power-law spectral models, the $\chi ^2$ and degree of freedom (dof) of the fit are given in the sixth column, and the seventh gives the null hypothesis probability.\label{tab:sedfits}}
\begin{tabular}{@{}ccccccc}
\hline
GRB & Model & \nhx & \av\ & $E_{bk}$ & $\chi ^2$ (dof) & Null Hypothesis\\
& & $10^{21}$~\invsqrcm & (mag) & (Hz) & & Probability\\
\hline\hline
050318 & SMC/pow & $1.40^{+0.42}_{-0.40}$ & $0.53^{+0.06}_{-0.06}$ &  - & 108 (88) & 0.068\\
& LMC/pow & $1.91^{+0.49}_{-0.47}$ & $0.78^{+0.09}_{-0.09}$ &  - & 113 (88) & 0.039\\
& MW/pow & $1.67^{+0.52}_{-0.49}$ & $0.88^{+0.13}_{-0.12}$ &  - & 147 (88) & 8.3e-05\\
& SMC/bknp & $1.41^{+0.43}_{-0.35}$ & $0.54^{+0.04}_{-0.02}$ &  8.616 & 108 (87) & 0.059\\
& LMC/bknp & $1.92^{+0.36}_{-0.34}$ & $0.79^{+0.09}_{-0.07}$ &  8.971 & 113 (87) & 0.034\\
& MW/bknp & $1.67^{+0.52}_{-0.49}$ & $0.88^{+0.12}_{-0.06}$ &  9.158 & 147 (87) & 6.2e-05\\
\hline
050319 & SMC/pow & $< 3.40$ & $0.07^{+0.04}_{-0.03}$ &  - & 81 (80) & 0.448\\
& LMC/pow & $< 4.36$ & $0.12^{+0.05}_{-0.05}$ &  - & 80 (80) & 0.492\\
& MW/pow & $< 5.25$ & $0.21^{+0.08}_{-0.07}$ &  - & 77 (80) & 0.588\\
& SMC/bknp & $< 4.16$ & $< 0.09$ &  2.541 & 71 (79) & 0.714\\
& LMC/bknp & $< 4.45$ & $< 0.21$ &  2.566 & 71 (79) & 0.722\\
& MW/bknp & $< 5.01$ & $< 0.35$ &  2.636 & 70 (79) & 0.755\\
\hline
050525A & SMC/pow & $2.25^{+0.41}_{-0.36}$ & $0.06\pm 0.01$ &  - & 50 (37) & 0.069\\
& LMC/pow & $2.24^{+0.41}_{-0.36}$ & $0.07\pm 0.02$ &  - & 56 (37) & 0.024\\
& MW/pow & $2.15^{+0.41}_{-0.36}$ & $0.06^{+0.02}_{-0.02}$ &  - & 68 (37) & 0.001\\
& SMC/bknp & $2.96^{+0.50}_{-0.48}$ & $0.16^{+0.02}_{-0.02}$ &  0.017 & 46 (36) & 0.119\\
& LMC/bknp & $3.12\pm 0.53$ & $0.21\pm 0.03$ &  0.022 & 57 (36) & 0.014\\
& MW/bknp & $3.82\pm 0.73$ & $0.10\pm 0.04$ &  0.724 & 84 (36) & 1.0e-05\\
\hline
050730 & SMC/pow & $14.88^{+2.26}_{-2.13}$ & $0.16^{+0.03}_{-0.02}$ &  - & 163 (133) & 0.040\\
& LMC/pow & $15.51^{+2.31}_{-2.18}$ & $0.22\pm 0.03$ &  - & 163 (133) & 0.039\\
& MW/pow & $16.20^{+2.38}_{-2.25}$ & $0.30\pm 0.05$ &  - & 164 (133) & 0.036\\
& SMC/bknp & $17.79^{+2.35}_{-2.10}$ & $0.23^{+0.02}_{-0.03}$ &  0.001 & 159 (132) & 0.055\\
& LMC/bknp & $18.31^{+2.48}_{-2.01}$ & $0.31^{+0.03}_{-0.05}$ &  0.001 & 159 (132) & 0.053\\
& MW/bknp & $18.14^{+2.45}_{-2.03}$ & $0.39^{+0.04}_{-0.06}$ &  0.001 & 160 (132) & 0.047\\
\hline
050802 & SMC/pow & $1.28^{+0.59}_{-0.55}$ & $0.06\pm 0.02$ &  - & 90 (69) & 0.043\\
& LMC/pow & $1.43^{+0.61}_{-0.56}$ & $0.10\pm 0.03$ &  - & 89 (69) & 0.056\\
& MW/pow & $1.74^{+0.64}_{-0.60}$ & $0.19\pm 0.06$ &  - & 84 (69) & 0.101\\
& SMC/bknp & $1.56^{+0.59}_{-0.58}$ & $0.05^{+0.01}_{-0.02}$ &  3.150 & 81 (68) & 0.138\\
& LMC/bknp & $1.67^{+0.59}_{-0.57}$ & $0.08^{+0.02}_{-0.03}$ &  3.168 & 80 (68) & 0.155\\
& MW/bknp & $1.89^{+0.65}_{-0.61}$ & $0.15^{+0.06}_{-0.02}$ &  3.233 & 77 (68) & 0.212\\
\hline
050820A & SMC/pow & $< 0.46$ & $0.18^{+0.01}_{-0.01}$ &  - & 224 (139) & 6.2e-06\\
& LMC/pow & $< 0.67$ & $0.29^{+0.03}_{-0.02}$ &  - & 208 (139) & 1.3e-04\\
& MW/pow & $< 1.45$ & $0.43^{+0.04}_{-0.04}$ &  - & 193 (139) & 0.002\\
& SMC/bknp & $5.07^{+1.25}_{-0.62}$ & $0.14\pm 0.03$ &  0.209 & 144 (138) & 0.350\\
& LMC/bknp & $5.04^{+1.26}_{-1.21}$ & $0.23\pm 0.04$ &  0.143 & 142 (138) & 0.386\\
& MW/bknp & $5.35^{+1.19}_{-1.21}$ & $0.32^{+0.06}_{-0.06}$ &  0.139 & 143 (138) & 0.365\\
\hline
050922C & SMC/pow & $1.62^{+0.89}_{-0.82}$ & $0.07\pm 0.02$ &  - & 36 (48) & 0.902\\
& LMC/pow & $1.73^{+0.90}_{-0.83}$ & $0.11^{+0.03}_{-0.03}$ &  - & 35 (48) & 0.915\\
& MW/pow & $1.85^{+0.92}_{-0.85}$ & $0.16^{+0.05}_{-0.05}$ &  - & 35 (48) & 0.925\\
& SMC/bknp & $3.03^{+0.97}_{-1.56}$ & $0.14^{+0.02}_{-0.02}$ &  0.005 & 34 (47) & 0.915\\
& LMC/bknp & $3.13^{+0.98}_{-1.35}$ & $0.21^{+0.04}_{-0.03}$ &  0.005 & 34 (47) & 0.926\\
& MW/bknp & $3.58^{+1.04}_{-1.31}$ & $0.28^{+0.06}_{-0.05}$ &  0.007 & 37 (47) & 0.851\\
\hline
\end{tabular}
\end{minipage}
\end{table*}

\begin{table*}
\centering
\begin{minipage}{126mm}
\setcounter{table}{3}
\caption{Continued}
\begin{tabular}{ccccccc}
\hline
GRB & Model & \nhx & \av\ & $E_{bk}$ & $\chi ^2$ (dof) & Null Hypothesis\\
& & $10^{21}$~\invsqrcm & (mag) & (Hz) & & Probability\\
\hline\hline
051109A & SMC/pow & $< 3.64$ & $< 0.14$ &  - & 239 (145) & 1.3e-06\\
& LMC/pow & $< 3.64$ & $< 0.19$ &  - & 239 (145) & 1.3e-06\\
& MW/pow & $< 3.64$ & $< 0.25$ &  - & 239 (145) & 1.3e-06\\
& SMC/bknp & $8.71^{+1.58}_{-1.46}$ & $< 0.10$ &  0.090 & 162 (144) & 0.141\\
& LMC/bknp & $8.75^{+1.59}_{-1.55}$ & $< 0.13$ &  0.092 & 162 (144) & 0.141\\
& MW/bknp & $8.80\pm 1.55$ & $< 0.17$ &  0.096 & 162 (144) & 0.140\\
\hline
060124 & SMC/pow & $2.60^{+0.87}_{-0.81}$ & $0.08\pm 0.03$ &  - & 182 (115) & 6.6e-05\\
& LMC/pow & $3.24^{+0.92}_{-0.86}$ & $0.17\pm 0.04$ &  - & 174 (115) & 3.0e-04\\
& MW/pow & $5.83^{+1.13}_{-1.05}$ & $0.56^{+0.09}_{-0.08}$ &  - & 144 (115) & 0.034\\
& SMC/bknp & $9.41^{+1.47}_{-1.37}$ & $0.13^{+0.03}_{-0.04}$ &  0.233 & 116 (114) & 0.439\\
& LMC/bknp & $9.39^{+1.45}_{-1.38}$ & $0.22\pm 0.06$ &  0.156 & 114 (114) & 0.483\\
& MW/bknp & $9.68^{+1.48}_{-1.31}$ & $0.52^{+0.13}_{-0.13}$ &  0.060 & 113 (114) & 0.498\\
\hline
060206 & SMC/pow & $< 6.72$ & $< 0.04$ &  - & 97 (53) & 2.0e-04\\
& LMC/pow & $< 6.72$ & $< 0.05$ &  - & 97 (53) & 2.0e-04\\
& MW/pow & $< 6.72$ & $< 0.05$ &  - & 97 (53) & 2.0e-04\\
& SMC/bknp & $13.65^{+3.38}_{-2.79}$ & $< 0.18$ &  0.618 & 60 (52) & 0.218\\
& LMC/bknp & $14.06^{+2.95}_{-2.92}$ & $< 0.05$ &  0.604 & 60 (52) & 0.218\\
& MW/bknp & $13.69^{+3.35}_{-2.83}$ & $< 0.22$ &  0.617 & 60 (52) & 0.218\\
\hline
060418 & SMC/pow & $3.60^{+1.83}_{-1.46}$ & $< 0.06$ &  - & 41 (24) & 0.017\\
& LMC/pow & $3.53^{+1.82}_{-1.46}$ & $< 0.09$ &  - & 42 (24) & 0.014\\
& MW/pow & $3.44^{+1.80}_{-1.44}$ & $< 0.10$ &  - & 42 (24) & 0.013\\
& SMC/bknp & $4.02^{+2.12}_{-1.34}$ & $0.09^{+0.01}_{-0.02}$ &  0.002 & 21 (23) & 0.591\\
& LMC/bknp & $4.24^{+2.27}_{-1.36}$ & $0.12^{+0.02}_{-0.03}$ &  0.001 & 23 (23) & 0.461\\
& MW/bknp & $3.65^{+3.63}_{-1.21}$ & $0.08^{+0.03}_{-0.02}$ &  0.001 & 29 (23) & 0.186\\
\hline
060502A & SMC/pow & $3.42^{+0.90}_{-0.80}$ & $0.51^{+0.12}_{-0.10}$ &  - & 33 (31) & 0.368\\
& LMC/pow & $4.03^{+0.96}_{-0.86}$ & $0.74^{+0.14}_{-0.13}$ &  - & 35 (31) & 0.299\\
& MW/pow & $3.88^{+1.03}_{-0.93}$ & $0.79^{+0.18}_{-0.17}$ &  - & 55 (31) & 0.005\\
& SMC/bknp & $5.07^{+1.24}_{-1.31}$ & $0.50^{+0.13}_{-0.10}$ &  0.031 & 30 (30) & 0.467\\
& LMC/bknp & $4.10^{+1.03}_{-0.86}$ & $0.76^{+0.14}_{-0.06}$ &  0.002 & 34 (30) & 0.297\\
& MW/bknp & $6.18^{+1.30}_{-1.48}$ & $0.50^{+0.20}_{-0.19}$ &  0.432 & 57 (30) & 0.002\\
\hline
060512 & SMC/pow & $< 0.85$ & $0.47\pm 0.05$ &  - & 84 (23) & 7.4e-09\\
& LMC/pow & $< 0.85$ & $0.56\pm 0.06$ &  - & 99 (23) & 2.1e-11\\
& MW/pow & $< 0.82$ & $0.67\pm 0.08$ &  - & 122 (23) & 2.0e-15\\
& SMC/bknp & $< 1.74$ & $0.66\pm 0.09$ &  0.007 & 77 (22) & 4.8e-08\\
& LMC/bknp & $< 1.79$ & $0.79\pm 0.13$ &  0.011 & 99 (22) & 1.0e-11\\
& MW/bknp & $< 2.03$ & $0.95\pm 0.15$ &  0.015 & 131 (22) & 2.1e-17\\
\hline
060526 & SMC/pow & $< 45.39$ & $< 0.07$ &  - & 9 (8) & 0.344\\
& LMC/pow & $< 45.61$ & $< 0.10$ &  - & 9 (8) & 0.344\\
& MW/pow & $< 45.67$ & $< 0.21$ &  - & 9 (8) & 0.345\\
& SMC/bknp & $< 47.14$ & $< 0.16$ &  0.002 & 7 (7) & 0.476\\
& LMC/bknp & $13.23^{+12.85}_{-7.60}$ & $0.10\pm 0.04$ &  0.002 & 6 (7) & 0.486\\
& MW/bknp & $14.69^{+11.18}_{-8.45}$ & $0.18^{+0.06}_{-0.09}$ &  0.002 & 6 (7) & 0.554\\
\hline
\end{tabular}
\end{minipage}
\end{table*}

\begin{table*}{}
\centering
\begin{minipage}{126mm}
\setcounter{table}{3}
\caption{Continued}
\begin{tabular}{@{}ccccccc}
\hline
GRB & Model & \nhx & \av\ & $E_{bk}$ & $\chi ^2$ (dof) & Null Hypothesis\\
& & $10^{21}$~\invsqrcm & (mag) & (Hz) & & Probability\\
\hline\hline
060605 & SMC/pow & $7.80^{+2.70}_{-2.58}$ & $0.25^{+0.06}_{-0.05}$ &  - & 62 (68) & 0.670\\
& LMC/pow & $7.79^{+2.71}_{-2.58}$ & $0.32^{+0.07}_{-0.07}$ &  - & 62 (68) & 0.673\\
& MW/pow & $6.47^{+2.54}_{-2.39}$ & $0.35^{+0.08}_{-0.08}$ &  - & 63 (68) & 0.662\\
& SMC/bknp & $7.70^{+2.73}_{-2.61}$ & $0.24^{+0.06}_{-0.06}$ &  4.892 & 62 (67) & 0.662\\
& LMC/bknp & $7.61^{+2.72}_{-2.65}$ & $0.31^{+0.07}_{-0.09}$ &  4.971 & 62 (67) & 0.664\\
& MW/bknp & $6.13^{+2.53}_{-2.44}$ & $0.31^{+0.10}_{-0.09}$ &  3.947 & 62 (67) & 0.666\\
\hline
060607A & SMC/pow & $6.19^{+1.79}_{-1.70}$ & $< 0.15$ &  - & 116 (117) & 0.506\\
& LMC/pow & $6.16^{+1.83}_{-1.73}$ & $< 0.21$ &  - & 116 (117) & 0.499\\
& MW/pow & $5.96^{+1.82}_{-1.72}$ & $< 0.26$ &  - & 117 (117) & 0.489\\
& SMC/bknp & $6.27^{+1.58}_{-1.69}$ & $< 0.11$ &  9.787 & 116 (116) & 0.479\\
& LMC/bknp & $6.22^{+1.61}_{-1.39}$ & $< 0.14$ &  9.786 & 116 (116) & 0.472\\
& MW/bknp & $6.04^{+1.61}_{-1.37}$ & $< 0.14$ &  9.832 & 117 (116) & 0.462\\
\hline
060714 & SMC/pow & $5.98^{+4.03}_{-3.62}$ & $0.46^{+0.17}_{-0.17}$ &  - & 20 (17) & 0.284\\
& LMC/pow & $6.48^{+4.41}_{-3.91}$ & $0.64^{+0.26}_{-0.25}$ &  - & 20 (17) & 0.268\\
& MW/pow & $< 20.04$ & $0.79^{+0.39}_{-0.35}$ &  - & 22 (17) & 0.196\\
& SMC/bknp & $< 17.07$ & $< 0.91$ &  2.341 & 19 (16) & 0.283\\
& LMC/bknp & $< 17.86$ & $< 1.28$ &  2.241 & 19 (16) & 0.261\\
& MW/bknp & $< 16.87$ & $< 1.66$ &  2.121 & 20 (16) & 0.211\\
\hline
060729 & SMC/pow & $0.80\pm 0.09$ & $0.03\pm 0.01$ &  - & 184 (178) & 0.367\\
& LMC/pow & $0.82\pm 0.09$ & $0.04\pm 0.02$ &  - & 183 (178) & 0.373\\
& MW/pow & $0.83^{+0.10}_{-0.09}$ & $0.06\pm 0.03$ &  - & 184 (178) & 0.354\\
& SMC/bknp & $1.09^{+0.13}_{-0.11}$ & $0.13^{+0.01}_{-0.01}$ &  0.006 & 176 (177) & 0.496\\
& LMC/bknp & $1.10^{+0.15}_{-0.09}$ & $0.18^{+0.03}_{-0.02}$ &  0.006 & 178 (177) & 0.465\\
& MW/bknp & $0.97^{+0.09}_{-0.12}$ & $0.15^{+0.03}_{-0.05}$ &  0.004 & 179 (177) & 0.453\\
\hline
060904B & SMC/pow & $1.84^{+0.41}_{-0.37}$ & $0.06\pm 0.02$ &  - & 83 (46) & 7.3e-04\\
& LMC/pow & $1.85^{+0.41}_{-0.37}$ & $0.08\pm 0.03$ &  - & 83 (46) & 6.3e-04\\
& MW/pow & $1.84^{+0.41}_{-0.37}$ & $0.09\pm 0.04$ &  - & 85 (46) & 3.7e-04\\
& SMC/bknp & $3.72^{+0.66}_{-0.76}$ & $0.12^{+0.05}_{-0.04}$ &  0.174 & 74 (45) & 0.004\\
& LMC/bknp & $3.65^{+0.74}_{-0.69}$ & $0.17\pm 0.06$ &  0.129 & 75 (45) & 0.003\\
& MW/bknp & $4.05^{+0.57}_{-0.68}$ & $0.14^{+0.08}_{-0.07}$ &  0.392 & 80 (45) & 0.001\\
\hline
060908 & SMC/pow & $< 8.74$ & $< 0.17$ &  - & 15 (8) & 0.061\\
& LMC/pow & $< 8.95$ & $< 0.38$ &  - & 15 (8) & 0.060\\
& MW/pow & $< 8.71$ & $< 0.22$ &  - & 15 (8) & 0.061\\
& SMC/bknp & $< 25.95$ & $< 0.21$ &  1.287 & 12 (7) & 0.111\\
& LMC/bknp & $< 13.77$ & $< 0.26$ &  1.212 & 12 (7) & 0.109\\
& MW/bknp & $< 25.82$ & $< 0.18$ &  1.288 & 12 (7) & 0.111\\
\hline
060912 & SMC/pow & $3.23^{+0.58}_{-0.52}$ & $0.44^{+0.12}_{-0.10}$ &  - & 60 (42) & 0.037\\
& LMC/pow & $3.52^{+0.65}_{-0.56}$ & $0.62^{+0.15}_{-0.14}$ &  - & 59 (42) & 0.040\\
& MW/pow & $3.74^{+0.72}_{-0.65}$ & $0.81^{+0.19}_{-0.18}$ &  - & 61 (42) & 0.027\\
& SMC/bknp & $3.23^{+0.59}_{-0.52}$ & $0.44^{+0.10}_{-0.11}$ &  5.975 & 60 (41) & 0.029\\
& LMC/bknp & $3.53^{+0.52}_{-0.46}$ & $0.62^{+0.06}_{-0.08}$ &  8.717 & 59 (41) & 0.031\\
& MW/bknp & $3.70^{+0.60}_{-0.42}$ & $0.82^{+0.07}_{-0.19}$ &  7.695 & 61 (41) & 0.021\\
\hline
\end{tabular}
\end{minipage}
\end{table*}

\begin{table*}{}
\centering
\begin{minipage}{126mm}
\setcounter{table}{3}
\caption{Continued}
\begin{tabular}{@{}ccccccc}
\hline
GRB & Model & \nhx & \av\ & $E_{bk}$ & $\chi ^2$ (dof) & Null Hypothesis\\
& & $10^{21}$~\invsqrcm & (mag) & (Hz) & & Probability\\
\hline\hline
061007 & SMC/pow & $4.58^{+0.19}_{-0.18}$ & $0.45\pm 0.01$ &  - & 380 (274) & 2.2e-05\\
& LMC/pow & $5.44^{+0.21}_{-0.20}$ & $0.75\pm 0.02$ &  - & 325 (274) & 0.019\\
& MW/pow & $6.82\pm 0.23$ & $1.26\pm 0.03$ &  - & 1157 (274) & 0.0e+00\\
& SMC/bknp & $5.10^{+0.19}_{-0.23}$ & $0.53^{+0.01}_{-0.01}$ &  0.004 & 337 (273) & 0.005\\
& LMC/bknp & $5.74\pm 0.21$ & $0.82^{+0.01}_{-0.03}$ &  0.003 & 315 (273) & 0.040\\
& MW/bknp & $6.79\pm 0.26$ & $1.26\pm 0.03$ &  9.992 & 1157 (273) & 0.0e+00\\
\hline
061121 & SMC/pow & $3.64^{+0.43}_{-0.40}$ & $0.34\pm 0.03$ &  - & 176 (131) & 0.005\\
& LMC/pow & $4.04^{+0.46}_{-0.43}$ & $0.49\pm 0.04$ &  - & 167 (131) & 0.018\\
& MW/pow & $4.46^{+0.50}_{-0.47}$ & $0.71\pm 0.06$ &  - & 163 (131) & 0.031\\
& SMC/bknp & $3.87^{+0.46}_{-0.44}$ & $0.28\pm 0.03$ &  2.738 & 139 (130) & 0.272\\
& LMC/bknp & $4.13^{+0.49}_{-0.46}$ & $0.40\pm 0.04$ &  2.829 & 138 (130) & 0.305\\
& MW/bknp & $4.32^{+0.52}_{-0.49}$ & $0.55\pm 0.07$ &  2.865 & 139 (130) & 0.270\\
\hline
061126 & SMC/pow & $3.14^{+0.42}_{-0.39}$ & $0.13^{+0.03}_{-0.03}$ &  - & 190 (133) & 8.9e-04\\
& LMC/pow & $3.35^{+0.44}_{-0.41}$ & $0.21^{+0.04}_{-0.04}$ &  - & 186 (133) & 0.002\\
& MW/pow & $3.50^{+0.46}_{-0.43}$ & $0.28^{+0.06}_{-0.06}$ &  - & 186 (133) & 0.002\\
& SMC/bknp & $5.95^{+0.72}_{-0.67}$ & $0.10\pm 0.04$ &  0.135 & 144 (132) & 0.225\\
& LMC/bknp & $5.97^{+0.68}_{-0.67}$ & $0.14^{+0.05}_{-0.05}$ &  0.122 & 145 (132) & 0.209\\
& MW/bknp & $6.10^{+0.73}_{-0.35}$ & $0.14^{+0.07}_{-0.07}$ &  0.170 & 149 (132) & 0.153\\
\hline
070110 & SMC/pow & $< 3.29$ & $0.29^{+0.06}_{-0.05}$ &  - & 55 (50) & 0.306\\
& LMC/pow & $< 4.23$ & $0.44^{+0.09}_{-0.08}$ &  - & 51 (50) & 0.432\\
& MW/pow & $< 5.34$ & $0.64^{+0.14}_{-0.12}$ &  - & 50 (50) & 0.472\\
& SMC/bknp & $< 3.48$ & $0.23^{+0.06}_{-0.05}$ &  2.408 & 45 (49) & 0.656\\
& LMC/bknp & $< 4.07$ & $0.34^{+0.09}_{-0.08}$ &  2.530 & 44 (49) & 0.692\\
& MW/bknp & $< 4.69$ & $0.49^{+0.14}_{-0.13}$ &  2.828 & 44 (49) & 0.662\\
\hline
070318 & SMC/pow & $8.52^{+0.84}_{-0.73}$ & $0.52\pm 0.02$ &  - & 69 (44) & 0.010\\
& LMC/pow & $8.96\pm 0.77$ & $0.73\pm 0.03$ &  - & 94 (44) & 1.5e-05\\
& MW/pow & $9.49\pm 0.81$ & $1.05\pm 0.04$ &  - & 213 (44) & 4.8e-24\\
& SMC/bknp & $8.78^{+1.02}_{-0.65}$ & $0.59^{+0.01}_{-0.06}$ &  0.003 & 65 (43) & 0.017\\
& LMC/bknp & $9.21\pm 1.14$ & $0.82\pm 0.05$ &  0.003 & 86 (43) & 1.1e-04\\
& MW/bknp & $9.21\pm 1.13$ & $0.82\pm 0.05$ &  0.003 & 86 (43) & 1.1e-04\\
\hline
070411 & SMC/pow & $< 32.05$ & $< 0.21$ &  - & 24 (17) & 0.117\\
& LMC/pow & $< 31.93$ & $< 0.30$ &  - & 24 (17) & 0.117\\
& MW/pow & $< 31.93$ & $< 0.47$ &  - & 24 (17) & 0.117\\
& SMC/bknp & $< 32.81$ & $< 0.20$ &  5.026 & 24 (16) & 0.089\\
& LMC/bknp & $< 32.68$ & $< 0.28$ &  5.026 & 24 (16) & 0.089\\
& MW/bknp & $< 32.71$ & $< 0.44$ &  5.026 & 24 (16) & 0.089\\
\hline
070529 & SMC/pow & $14.12^{+6.51}_{-5.32}$ & $< 0.52$ &  - & 20 (25) & 0.728\\
& LMC/pow & $14.23^{+6.58}_{-5.37}$ & $< 0.68$ &  - & 20 (25) & 0.729\\
& MW/pow & $13.76^{+6.36}_{-5.19}$ & $< 0.84$ &  - & 20 (25) & 0.724\\
& SMC/bknp & $13.52^{+6.39}_{-5.26}$ & $< 0.45$ &  3.734 & 18 (24) & 0.780\\
& LMC/bknp & $13.61^{+6.46}_{-5.32}$ & $< 0.58$ &  3.698 & 18 (24) & 0.780\\
& MW/bknp & $13.41^{+6.21}_{-5.12}$ & $< 0.70$ &  3.713 & 18 (24) & 0.781\\
\hline
\end{tabular}
\end{minipage}
\end{table*}

\end{document}